\documentclass[preprint,aps]{revtex4}

\usepackage{graphicx}
\usepackage{tensind}
\tensordelimiter{?}
\usepackage{doublespace}

\usepackage{amssymb}
\usepackage{amsmath}

\usepackage{booktabs} 

\usepackage{dsfont}

\usepackage{booktabs} 

\usepackage{hyperref}

\def\la{\lambda}
\def\w{\omega}

\def\a{\alpha}
\def\be{\beta}

\renewcommand\th{\theta}

\def\*{\cdot}

\def\De{\Delta}
\def\d{\partial}

\def\de{\delta}

\def\ba#1{\overline{#1}}
\def\v#1{\mathbf{#1}}
\def\u#1{\underline{#1}}

\def\t{\tilde}
\def\vp{\varphi}
\def\Om{\Omega}

\def\ka{\kappa}

\newcommand\ra{\rightarrow}

\def\jd{\tfrac{1}{2}}

\newcommand\Rb{$^{87}$Rb}
\newcommand\axun{e_{\Phi}}

\newcommand\cm{\text{cm}}
\newcommand\sek{\text{s}}

\newcommand\meter{\text{m}}

\newcommand\sgn{\text{ sgn\,}}
\newcommand\He{\text{He}}
\newcommand\TT{\mathcal T}
\newcommand\HH{\mathcal K}
\newcommand\KK{\mathcal H}
\newcommand\RR{\mathbb R}
\newcommand\PP{\mathfrak p}

\newcommand\mm{\kappa}

\def\vec#1{\mathbf #1}
\def\pers#1{{#1}}

\def\Bessel#1#2{\text{Bessel}_{#1}(#2)}
\def\Be{\text{Bessel}}

\def\CC{\mathcal C}
\def\SS{\mathcal S}
\def\ZZ{\mathbb Z}
\def\CCC{\mathbb C}

\def\cauchy{{\CC_t}}

\def\g{\gamma}

\def\s{\sigma}

\newcommand\eqdef{\stackrel{\rm def}{=}}

\def\Fulling1{PhysRevD.14.1939}
\def\citCase{PhysRev.80.797}
\def\StoneBA{PhysRevB.61.11780}
\def\StoneWaveEq{PerezBergliaffa2004121}

\begin{document}

\title{Propagation of sound on line vortices is superfluids: role of ergoregions}
\author{Piotr Marecki\\ {\it Fakult\"at f\"ur Physik, Universit\"at Duisburg-Essen, Lotharstrasse 1, 47057 Duisburg, Germany
}}

\begin{abstract}
We (re)cosider the propagation of small disturbances (sound waves) in the presence of a pinned irrotational vortex in a superfluid with the help of the formalism of acoustic spacetimes. We give closed formulas for the scattering angle for sound rays, formulate the sound-propagation problem in the Hamiltonian form, and discuss the form of boundary conditions at the core of the vortex, where the Hamiltonian has a singular point. The wave equation is simplified to a single ordinary differential equation of Mathieu type. We give an extensive discussion of perturbations localized close to the core, which are similar to what is known as the Kelvin waves. The spectra of modes depend strongly on the type of boundary condition employed close to the vortex core. The existence of the gapless Kelvin mode, which is one of the modes with angular number -1, is usually discussed in the context of unpinned vortices in superfluid helium or rotating Bose-Einstein condensates. We prove that this particular mode is absent if the vortex is pinned, and consequently one must discuss the full family of modes in this case. The question of whether or not the acoustic spacetime admits an ergoregion turns out to have a decisive (qualitative) influence on many aspects of sound-propagation phenomena. 

\end{abstract}

 \maketitle
\setcounter{section}{0}

\section{Introduction}

\subsection{Background}
Vortices provide quantum fluids with a way to carry angular momentum \cite{Onsager,Feynman_helium}. In the simplest case of a vortex  line the stationary velocity field, $\vec v(\vec x)$, is purely azimuthal and depends as $|\vec v|=const/r$ on the distance $r$ from the line. This flow is locally irrotational but nonetheless possesses a finite global circulation $\Gamma=\int \vec v \cdot d \vec x$. For fluids with phase coherence $\Gamma$ must be an integer multiple of $h/m_B$, where $m_B$ is the mass of the bosonic constituent of the fluid, e.g.  the mass of a single helium-4 atom for the superfluid helium-4. A way to probe rotating superfluids is provided by sound waves, \cite{Donnelly}. In investigations of sound choices need to be made with regard to the type of description of the fluid and the structure of the vortex core (the innermost part of the vortex). Initial research on this topic was mainly concerned with the superfluid helium-4 described by equations of fluid-dynamics \cite{Fetter_sound,Donnelly}. Due to the growing experimental accessibility of quantum vapors (BECs)  the recent works tend to deal with this type of fluid, employing various approaches stemming from the Bogoliubov theory, \cite{Isoshima_main,Isoshima,BretinPhD}. The correspondence to earlier research is provided by (but also limited to the range of applicability of) the relation between these detailed descriptions and the fluid dynamics.   

In the case of experiments with vortices in helium the typical configurations consist of cylindrical cryostats filled with the superfluid, with the possibility of placing a wire (or rod) on the axis of symmetry of the cylinder. For not too large angular momenta of the fluid the velocity field corresponding to the stationary flow is exactly the one of the vortex line, while for higher angular momenta a vortex lattice is formed \cite{Donnelly}. The interesting problem of finding the most stable stationary state of the fluid for given radii of the cylinders at a given temperature and angular momentum of the fluid has been intensively investigated by many authors \cite{Fetter_annulus,Baym,Sonin}.

 On the other hand, in the case of BECs, the typical configurations consist of trapped rotating clouds of supersaturated vapors. For repulsively interacting constituents the vapor settles down to a metastable state, which is well described as a weakly-interacting Bose gas \cite{PethickSmith,Stringari}. The type of the most stable configuration of a trapped gas with a given angular momentum depends strongly on the form of the trapping potential (the potential seen by the constituents in the plane perpendicular to the  axis of rotation): for harmonic trapping an arrangement of single vortices (each carrying a single quantum of circulation) is the stable configurations for not too large angular momenta. Harmonic trapping has the special feature, that above certain angular momentum the BEC cloud begins to shed atoms radially. However, for trapping potentials ``steeper'' at large distances from the axis larger angular momenta can be accommodated by the cloud. When this is done, a configuration with a giant vortex in the middle of the trap is the expected final configuration at large angular momenta \cite{Baym}. A single vortex (simple or giant) provides the velocity field of a line vortex.\\

	The axis of rotation of an idealized vortex line is a singular place because $|\vec v|=const/r$ cannot continue to $r=0$. Dynamical models of vortex cores regularize this singularity. On the level of fluid dynamics one may distinguish two typical types of models: for superfluids, if an equation of state is given (and the symmetry of the flow is assumed) then we are lead to a vortex model with an interior surface there, where the pressure reaches the lowest allowed value (the  saturated vapor pressure). In this way an irrotational everywhere regular velocity field is obtained. Alternatively, for normal fluids, one assumes (as in the e.g. Rankine- \cite{FischerFlaig}, or Lamb-Oseen \cite{Fabre1} models) that there is a distribution of vorticity mostly localized in the innermost part of the vortex. In this way the irrotational $const/r$ profile is not continued towards the core, but rather the fluid velocity remains finite everywhere. Such vortex-core models are apparently adequate for the description of real phenomena in aerodynamics, \cite{Fabre1}. On the other hand, in the case of rotating quantum vapors (BECs) there exist natural descriptions of the innermost region, alternative to fluid dynamics, \cite{Isoshima_main,PethickSmith,Stringari}, perhaps the simplest of which being the Gross-Pitaevskii description. While in regions where the condensate density is large and slowly varying in space the BEC can be described as a (super)fluid, this is not anymore the case in the core. Even though the equations of the fluid dynamics fail there, the mean-field Gross-Pitaevski (or other) description is still valid, and essential characteristics of excitations of the vortex can be computed reliably (numerically), see e.g. \cite{Isoshima_main,Isoshima}. While these characteristics are very important for experimental verifications, they are nonetheless specific to the case of weakly interacting quantum vapors. In this paper we chose to work within the universality class provided by equations of fluid dynamics. We show, that qualitative difference is expected for sound propagation phenomena once the region to the interior of $r_c=\Gamma/c_s$ (with $c_s$ denoting the local speed of sound) is accessible to the fluid.

	In the variety of configurations outlined above the problem of sound propagation has been considered by many authors. The simplest approach, also used in this work, is to employ the macroscopic, fluid-dynamic description. The problem is further simplified if the perturbation of density and velocity can be assumed small (linearization). Care must be exercised with this assumption if the position of the vortex is allowed to move as a result of the perturbation. If the velocity the flow to be perturbed is large close to the core (i.e. if the modification of $const/r$ profile is localized at small $r$) then even a small shift of the axis of rotation may mean a large perturbation of the velocity field. In this work we consider only pinned vortices, for which the above obstruction does not occur. The other way to address the above problem is to consider regularized core models \cite{Saffman,Sipp2003,Fabre1} with sufficiently large cores, so that the background velocity is not too large anywhere. The study of perturbations naturally splits into the consideration of \emph{sound scattering} and of \emph{perturbations bound to the vortex}.  
	
	For the \emph{scattering} the natural problem is to derive the phase shifts in the partial-wave expansion of perturbations. Approximate approaches to the problem start with the work of \pers{Fetter} \cite{Fetter_sound}, and were continued with growing sophistication. Some renewed activity in the field appeared in the last decade due to the connection to the Iordanskii force problem, \cite{Sonin2,PhysRevB.61.11780,FischerFlaig}, and the controversy associated with it. We remark, that the general consensus on independence of scattering characteristics (for small frequencies) from the details of the vortex core can easily be shown unfounded in the case of a pinned vortex admitting an ergoregion. We do not focus on this problem, however, in the present work. 
	
	As to the study of \emph{disturbances localized close to the vortex core} it is important to note that generally - for each model of the vortex core - there exists a large family of such. The characteristics of the allowed modes, such as the dispersion relations for the (many) branches existing for each angular number $m$, depend strongly on the type of regularization of the vortex core \cite{Saffman,Sipp2003,Fabre1}. This family is usually referred to as the family of \emph{Kelvin waves}. In many cases, there exists an important mode of vibrations with the angular number $m=-1$, positive frequency $\w$, and the gapless dispersion relation $\w(p)\approx p^2\ln(1/p)$ for small wavenumbers $p$ (wavenumber along the vortex). This mode is usually referred to as \emph{the Kelvin mode}, \cite{Donnelly,Fetter_Review}. Modes with $|m|=1$ (there are many such modes for a given $p$, \cite{Saffman,Sipp2003}) generally put the cores of unpinned vortices into motion. This has lead many authors to seek for simple explanations of these modes on the phenomenological level (e.g. by considering forces acting on the core, as if it were a string with a specific tension, \cite{Donnelly,Sonin}). It should be remarked, that a version of the Kelvin \emph{mode} has also been apparently found in BEC systems \cite{Fetter_Review,Isoshima}, although it is puzzling why none of the other modes of the family (which is usually very rich) has ever been reported to be found in this context. 
	
	Because for vortices pinned on sound-reflecting wires a qualitatively different dynamics in the innermost part is at work, and because such vortices do not allow for deformations of the core (vortex line cannot be considered as a deformable string), we shall refer to the modes localized close to the core/wire discussed in this paper as the \emph{quasi Kelvin waves} (or whispering-gallery modes, \cite{PM_RS}). This is further justified by the fact that, as we shall show, the familiar \emph{Kelvin mode}, with its characteristic gapless dispersion relation, is absent in the system for two types of one-parameter boundary conditions we consider.\\

	The main purpose of this paper is to report on new results for sound propagation in the presence of a pinned vortex line. We use the fluid-mechanical description and employ the formalism of acoustic spacetimes, which has been under intensive development recently, \cite{Visser_review}. In this way the mathematical structure of the problem receives additional attention. The sound fields are parameterized by a single scalar potential, and a simple Hamiltonian structure is available \cite{PerezBergliaffa2004121}. Quite remarkably, many technical aspects of the problem turn out to be related to other important areas of theoretical physics, including: stability of (classical and quantum) fields in the presence of ergoregions, behavior of fields in the presence of strongly binding potentials, and the Aharonov-Bohm scattering. While the last relation is well known among the authors \cite{Sonin2,PhysRevB.61.11780}, we point its limits once the acoustic spacetime contains an ergoregion. To the best of our knowledge the following results are new: exact formula for the scattering angle in ray-acoustics approximation (section \ref{geodesics}), discussion of the general form of boundary conditions at the core (``core conditions'', section \ref{section_boundary_condition}), (semi-numerical) determination of spectra of bound states (quasi-Kelvin/whispering-gallery waves) localized close to the vortex for Neumann and ``core'' boundary conditions (section \ref{bound_section}), and the discussion of various sections of the parameter space displaying the characteristics of these states (section \ref{bound_dispersion}). The recurrent theme, throughout the paper, is the the qualitative change of the character of the problem once the acoustic spacetime contains ergoregions. Some of the problems related to stability of the flow are formulated, but left to a future investigation (see section \ref{vortex_stability}).
	
The paper should be regarded as an attempt to strengthen the understanding of the mathematical structure of the sound-propagation problem for perturbations of  the vortex flow. The additional features reported here in the fluid-dynamical approach (e.g. the specific characteristics of the full family of quasi-Kelvin waves) may or may not find their counterparts in the more sophisticated descriptions of concrete superfluids, such as the Bogoliubov descriptions of weakly-interacting rotating Bose-Einstein condensates \cite{Isoshima,BretinPhD}. In section \ref{experiments} we present a short discussion of the possible experimental regimes, where these features might be detectible. The concrete case of superfluid helium-4 is considered in a greater detail elsewhere \cite{PM_RS}. In our opinion, problems of superfluid rotation are of such importance that progress achieved even in very simplified models may turn out very valuable for future research.

\section{Acoustic ray description}\label{geodesics}
In this paper we investigate propagation of sound in the presence of an irrotational flow of a vortex line. The vortex is assumed to be pinned (no deformations of the line allowed), and the density perturbation is assumed to be ``small'' (linear regime). The propagation of sound for such a flow can be derived from a single partial differential equation for a single scalar potential, $\psi$. The velocity- and density-perturbations are expressed by, \cite{\StoneWaveEq,Barcelo}, 
\begin{align}
\de v_i &=\d_i\psi,  \\
\de \rho &=-\frac{\rho}{c_s^2}\, (\d_t\psi+v^i\d_i\psi),
\end{align} 
where $c_s$ denotes the local speed of sound, $v$ the local velocity field, and $\rho$ the local mass-density of the background fluid configuration. The potential $\psi$ fulfills the d'Alembert equation associated with the so-called acoustic metric, the form of which is
\begin{equation}
ds^2=\frac{\rho}{c_s} \left[-c^2_sdt^2+\de_{ij}(dx^i-v^i\, dt)(dx^j-v^j\, dt)\right],
\end{equation}
 In the stationary, cylindrically-symmetric case of an irrotational vortex line it reduces to
\begin{equation}\label{spacetime}
ds^2=\frac{\rho}{c_s}  \left[-f dt^2-2\mm\, dtd\vp+r^2 d\vp^2+dr^2+dz^2\right], 
\end{equation}
where $f=c^2_s-\mm^2/r^2$ and $\mm=v_\vp=\tfrac{\Gamma}{2\pi}=\tfrac{1}{2\pi}\int \vec v d\vec{x}$ is a constant of dimension $[\mm]$=[{cm$^2$/s]. Positive $\mm$ corresponds to a fluid rotating in the positive direction. This constant, together with a reference value of $c_s$, e.g. at $\infty$, allows us to construct a length scale, $r_c=\mm/c_s$, and a time scale $t_c=\mm/c_s^2$. In what follows we will use dimensionless quantities $t/t_c$ and $r/r_c$, denoted by the same letters $(t,r)$ for brevity. From now on we will make the simplifying assumption, that $c_s$ and $\rho$ do not depend on the distance $r$ from the axis. Whether or not this is a restrictive assumption depends on the type of superfluid considered. We discuss the deviations from this assumption for superfluid helium-4 in chapter \ref{experiments}. The case with variable $\rho$ and $c_s$ is considered in a separate work, \cite{PM_RS}. With the assumption, and the rescaling, we are lead to considering physics of scalar fields in the spacetime with the metric
\begin{equation}\label{s_spacetime}
ds^2=-(1-\tfrac{1}{r^2})dt^2 -2dt d\vp +r^2 d\vp^2+dr^2+dz^2.
\end{equation}
The ranges of coordinates are standard, save for $r$, which in some cases will be restricted from below by $r_0$.

The first task is to determine null geodesics, corresponding to the acoustic rays (ray acoustics approximation). The equations of motion can be derived from the Lagrangian
\begin{equation}
L=\tfrac{1}{2}\, (-f\dot t^2-2\dot t\dot \vp +r^2 \dot \vp^2 +\dot r^2 +\dot z^2)=\tfrac{1}{2}\, \dot x_a \dot x^a=0,
\end{equation}
where $f=1- \tfrac{1}{r^2}$, and where differentiations are performed with respect the parameter $\tau$ of the geodesics, as $x^a=x^a(\tau)$. The last equity is due to the null character of the rays. The spacetime contains an ergoregion, if points with $r<1$ are accessible. This may, or may not, be the case, depending on the experimental configuration. If the vortex is pinned on a central wire, then the (rescaled) radius of the wire provides the lowest value of $r$, the $r_0$. The problem of integrating the geodesic equations reduces to quadratures, because there are four first integrals: the momentum along the vortex,  $p=\dot z$, the ``angular momentum'' $J$, the ``energy'' $E$, 
\begin{align}
J&=r^2\dot \vp - \dot t ,\\
E&=f\dot t + \dot \vp,
\end{align}
together with the Lagrangian $L$. Note that objects moving along the the zero-angular-momentum curves, $J=0$, possess the (coordinate) angular velocity $\tfrac{d\vp}{dt}=\tfrac{1}{r^2}$, and thus remain at rest with respect to the fluid particles (i.e. co-rotate with them).

 With the help of $E$, $J$, and the Lagrangian we obtain the identity $\tfrac{1}{2}(-E\dot t +J \dot \vp+\dot r^2+\dot z^2)=0$, which is fulfilled along the geodesic. The reparametrization invariance of null geodesics allows us to set $J$ or $E$ to any value; we chose $E=1$, leading to $\dot t=1$ at $r=\infty$ (provided, geodesics reach that point).  The velocities $(\dot t,\dot \vp)$ are eliminated by
\begin{align}
\dot t &=r^2\dot \vp-J, \\
\dot \vp &=\frac{1+fJ }{r^2}\ \Rightarrow\
\dot t =1-\frac{J}{r^2}.  \label{dotphi}  
\end{align}

(The value of $t$ could, as it stands, decrease\footnote{The ``t'' coordinate in the acoustic spacetime models has a special role, because it coincides with flow of time in the laboratory frame, so that events in the acoustic spacetime can be directly interpreted. On this ground it is evident that no ray propagating initially with $\dot t >0$ can switch to $\dot t <0$, at least as long as the ray acoustics is valid.} along a null geodesic with $J>0$ if this geodesic, coming from $r=\infty$, crossed the radius $r^2_{caus}=J$. This does not happen, however, as we shall see below.)

The only equation to solve is the radial equation, which now assumes the form 
\begin{equation}
 \dot r^2=\left(1-\frac{J}{r^2} \right)(1+fJ)-J-p^2.   
\end{equation}

\subsection{General properties of rays}
Let us consider rays which reach $r=\infty$ and call them scattering rays. The limit $r\ra\infty$ of the radial equation and of Eq. \eqref{dotphi} yields 
\begin{equation}
\dot r^2_\infty+p^2=1, \qquad \dot \vp=\frac{1+J}{r^2}. 
\end{equation} 
Consequently, it is possible and advantageous to introduce the standard scattering parameters. 
All scattering rays are specified by: the projection of the velocity on the radial direction, 
\begin{equation}
v\eqdef  \dot r_\infty=\cos(\a)
\end{equation}
and by the ``impact parameter'', $b$, defined in the way familiar from scattering theory
\begin{equation}
\dot \vp=\frac{b\, v}{r^2}, \qquad  b\, v=1+J,
\end{equation}
(at $r=\infty$). We obtain the following form of the classical scattering problem:
\begin{equation}\label{rdot_o}
\dot r^2=\left[1-\frac{J}{r^2}\, \right]\left[J+1-\frac{J}{r^2}\, \right]-(J+1)+v^2.
\end{equation}

Let us mention, that the second type of rays appearing is obtained by taking $|p|>E=1$, that is with $v^2<0$. These rays do not reach $r=\infty$ and therefore are \emph{rays bound to the vortex}. Setting $\mathcal E=v^2$ (and allowing for negative $\mathcal E$) we investigate the polynomial in $\tfrac{1}{r}$ on the right hand side of 
\begin{equation}\label{rdot}
\dot r^2=J^2 (\tfrac{1}{r})^4-J(J+2)(\tfrac{1}{r})^2+\mathcal E,
\end{equation}
which is just a rearrangement of \eqref{rdot_o}. Zeros of this polynomial correspond to peri-/apocenters of the rays, and motion can only take place in regions where the polynomial is positive. For $\mathcal E>0$, motion will always be possible at the two limits: $r\ra \infty$ and $r\ra 0$. Setting $\tfrac{1}{r^2}=w$ we note, that the product of the zeros is equal to $\mathcal E/J^2$. For the bound rays, with $\mathcal E<0$, exactly one zero of the polynomial is always positive, say ($w_+$), and it corresponds to exactly one  positive root for $\tfrac{1}{r}$.  Therefore, the bound rays are coming from $r=0$, reach the apocenter, and fall towards $r=0$. The case of scattering rays, with $\mathcal E>0$, requires a closer look.

\subsection{Determination of orbits for scattering rays}
In this section we consider the scattering rays, and determine the form of their orbits, i.e. the functional dependence of $1/r=u$ on the angle $\vp$. Quite generally we shall see, that geodesics starting from $r=\infty$ either reach a pericenter (and scatter back to $\infty$) or propagate towards and end on the vortex core. The problem is integrable in terms of elliptic functions. The range of angular momentum $J$ leading to absorption is, as expected, asymmetric due to the rotation of the vortex.

We start with 
\begin{equation}
\dot r=\frac{du}{d\vp}\,\frac{dr}{du}\,\frac{d\vp}{d\tau}\,   =-u'[J+1-J u^2]
\end{equation}
 and find from \eqref{rdot}
\begin{equation}
 (u')^2\, [J+1-Ju^2]^2=[J+1-Ju^2][1-Ju^2]-(J+1) +v^2,
\end{equation}
Thus our problem is the classical Jacobi's inversion problem for $u(\vp)$ on a Riemann surface of genus one, 
\begin{equation}\label{main_integral}
\int_{u(\vp)}^{u_0}\frac{du\, [(J+1)-Ju^2]}{\sqrt{P_4(u)}}=\vp,
\end{equation}
with 
\begin{equation}
P_4(u)=J^2u^4-J(J+2)u^2+v^2.
\end{equation}
In order to compute the scattering angle $\De$ we set $u_0$ to the lowest positive root of $P_4$ (pericenter), where we can assume $\vp=0$, and compute the integral up to $u(\De)=0$ (corresponding to $r\ra\infty$). The total scattering angle, $\th$,  will be $\th=\pi-2\De$. Before establishing the dependence of $\De(v,b)$ let us first consider the case when no scattering takes place, because there is no pericenter, an the ray propagates to the vortex core.

\subsubsection*{Fall onto the core}
The polynomial $P_4(u)$ is bi-quadratic. Let $w=u^2$. Both roots of $P_2(w)=P_4(\sqrt{w})$ have the same sign due to $w_+w_-=v^2/J^2$. In two ranges of parameters $(J,v)$ none of the roots, $w_\pm$, is real and positive (i.e. they do not lead to real values of $u$, and therefore the ray continues $r=0$): they are not real unless $(J+2)^2-4v^2\geq 0$, and they are real but negative for $J\in[2v-2,0]$. These ranges of parameters have been depicted in the Fig. \ref{abs_ka_v}. For other values of $(J,v)$ there are two real, positive roots, and thus four real roots of $P_4(u)$, two of which are positive, say $u_\pm$. The rays starting at $r=\infty$ get reflected at $u_-$, while the ones starting at $r=0$ get reflected at $u_+$. 
\begin{figure}[htb]
\centering
\includegraphics[scale=0.63]{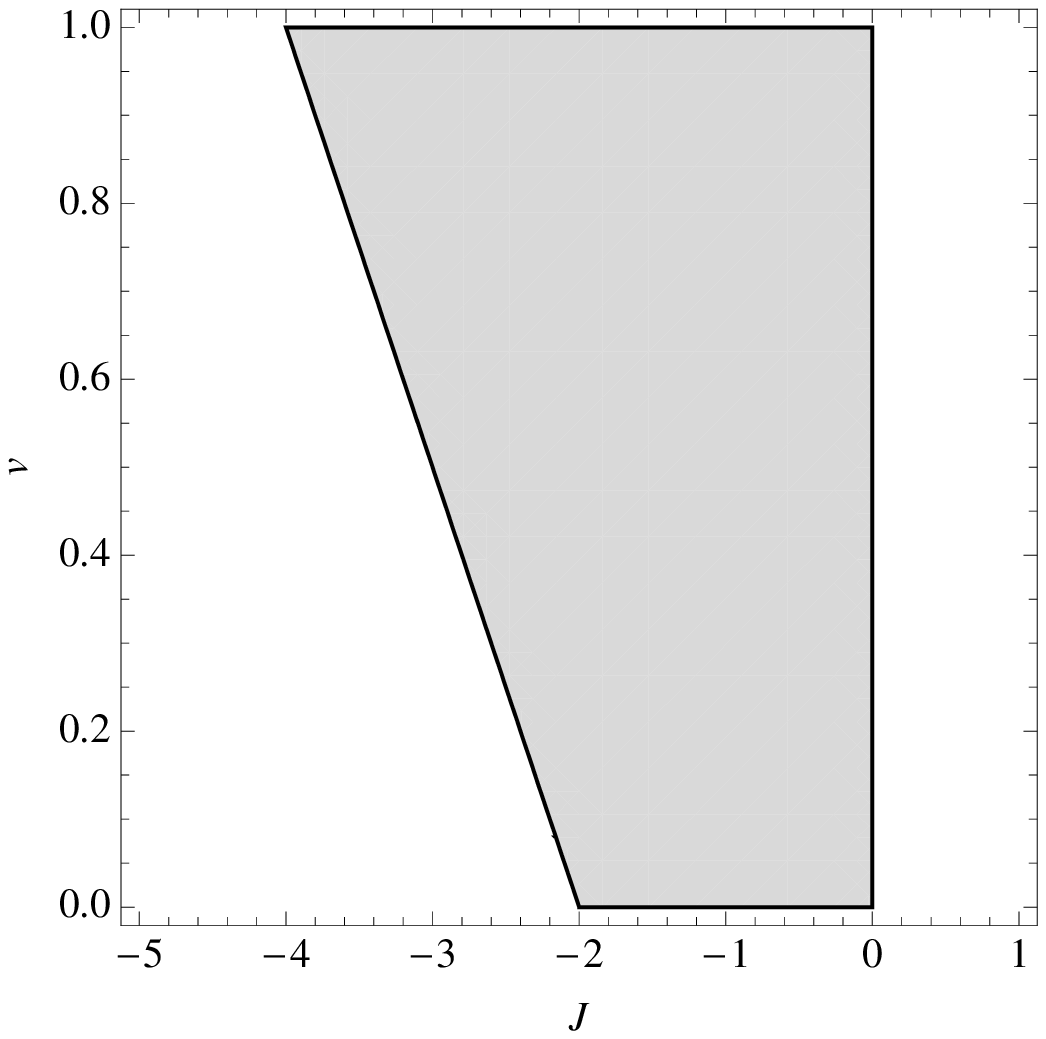}\quad\quad
\includegraphics[scale=0.63]{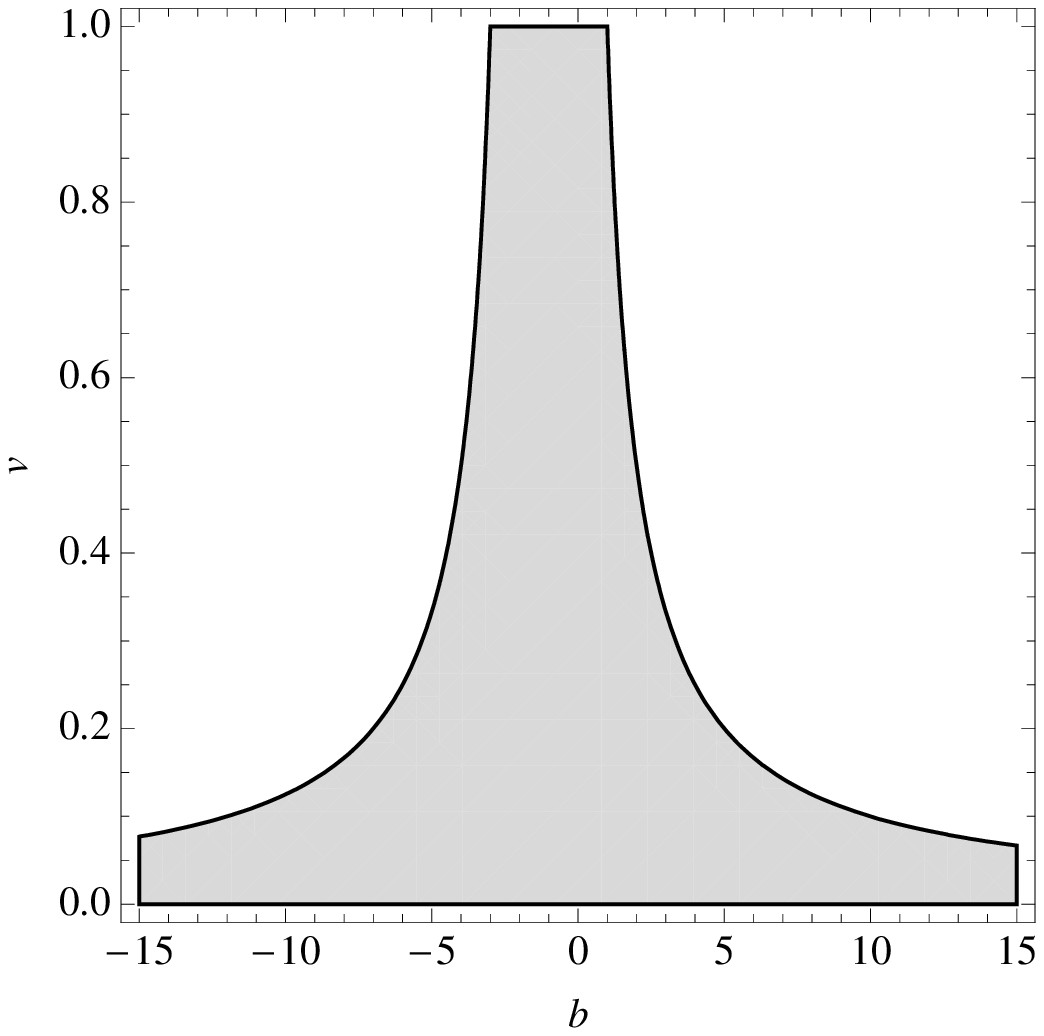}
\caption{The range of parameters for capture of rays coming from $\infty$: [Left:] $(J,v)$ plane [Right:] $(b,v)$ plane, with $bv=J+1$. For small $v$, that is for rays with small velocities $\dot r_\infty=v$ (propagating along the vortex line) the capture occurs for almost all values of the impact parameter $b$.}
\label{abs_ka_v}
\end{figure}

By considering the limiting form of the radial and axial equations we find the following asymptotic behavior of rays falling onto $r=0$:
\begin{equation}
\dot r^2= \frac{J^2}{r^4}, \qquad \dot \vp= -\frac{J}{r^4},
\end{equation}
from which it turns out, that the asymptotic trajectory is an (independent of $J$) hyperbolic spiral:
\begin{equation}
-\frac{dr}{r^2}=d\vp\ \Rightarrow \vp=\vp_0+\frac{1}{r}.  
\end{equation}

\begin{figure}[htb]
\centering
\includegraphics[scale=0.63]{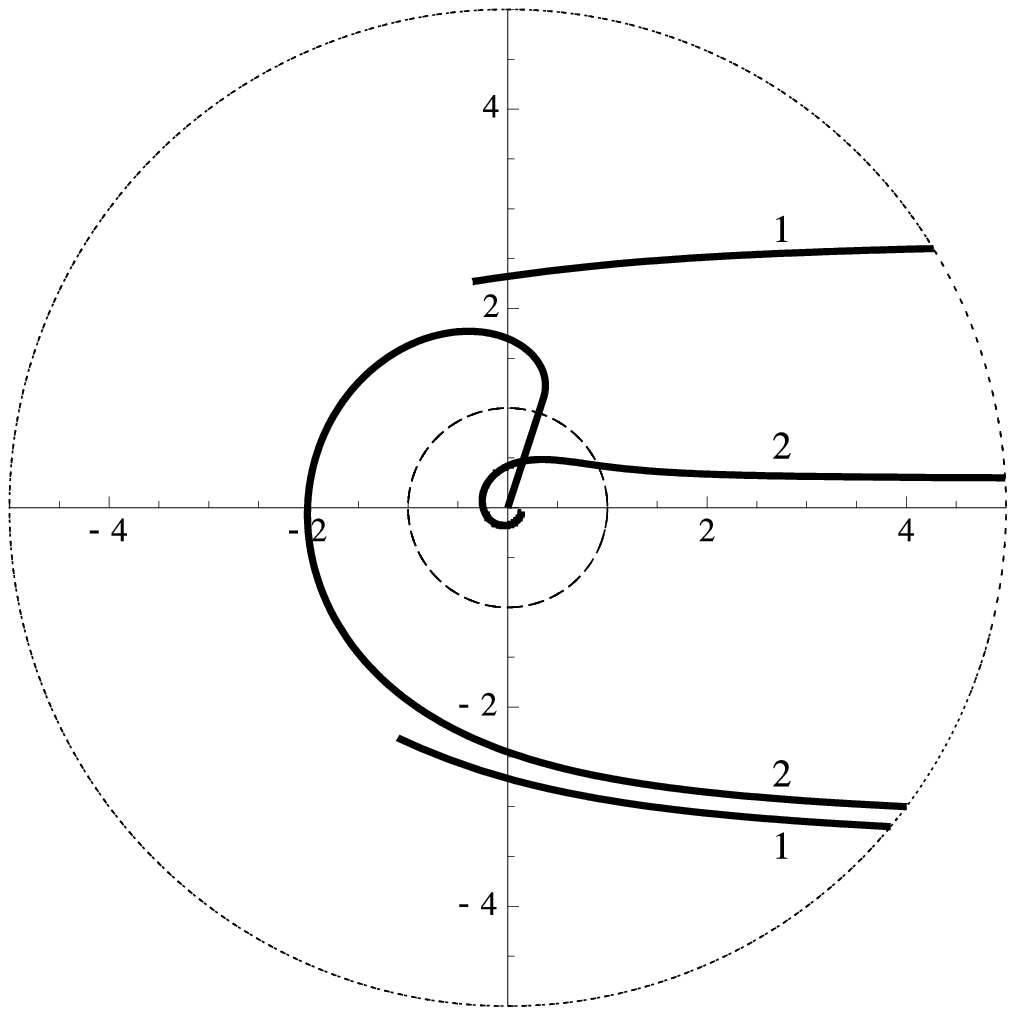}\quad
\includegraphics[scale=0.63]{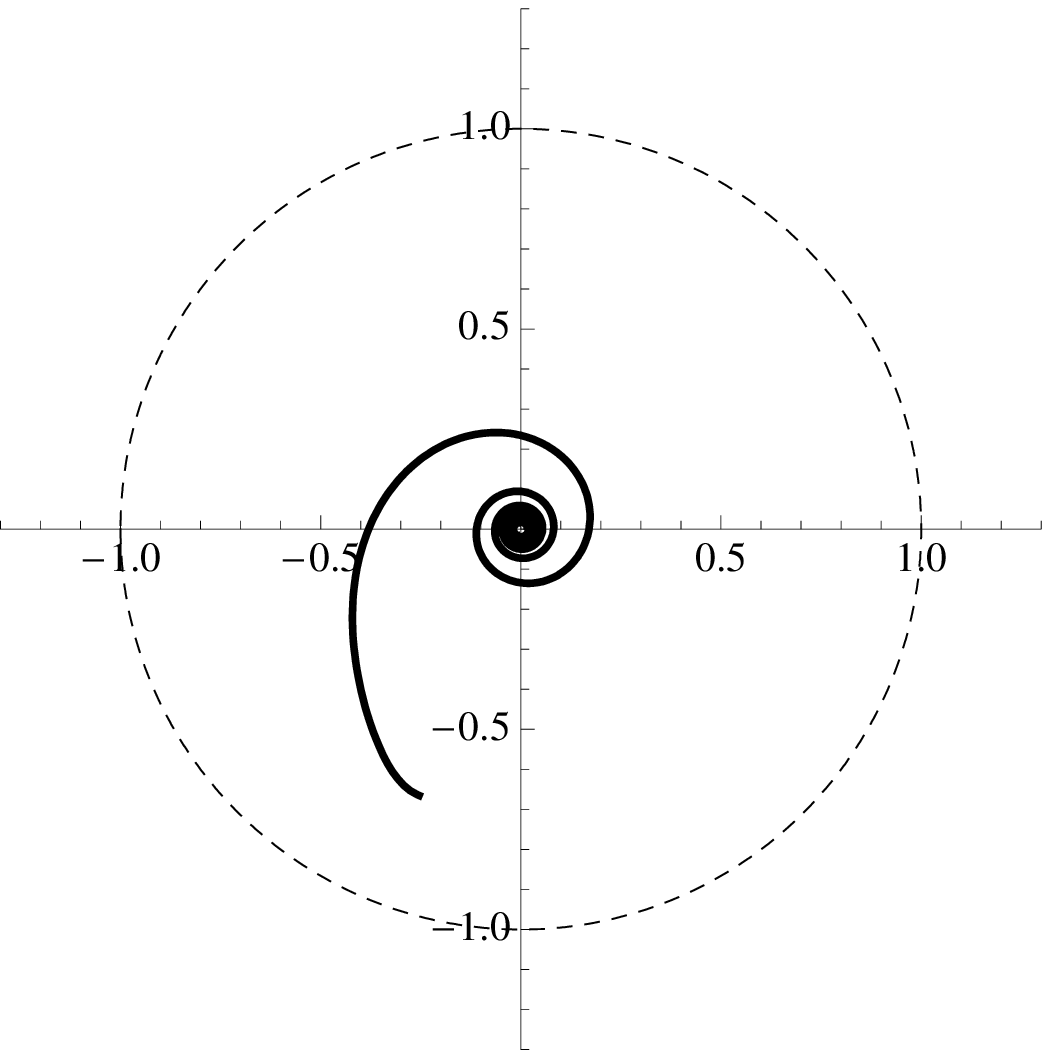}
\caption{Exact (numerical) solutions for sound rays. [Left:] Acoustic rays coming from $r=\infty$, which (1) get scattered by the vortex, (2) get absorbed by the vortex. [Right:] Rays emitted from the core which get reabsorbed. Paths have been plotted only up to the peri-/apocenter, where appropriate. Also visible is the boundary of the ergoregion, located at $r=1$. The flow in the vortex is positive-oriented.}
\label{geod}
\end{figure}

\subsubsection*{Exact results for the scattering angle $\De$}
Appearance of a quartic in the radical indicates a solution in terms of complete elliptic integrals. We recall the definitions, \cite{Abramowitz}:
\begin{equation}
K(m)=\int_0^1\frac{dt}{\sqrt{(1-t^2)(1-mt^2)}}, \qquad E(m)=\int_0^1\sqrt{\frac{1-mt^2}{1-t^2}\, }\, dt, 
\end{equation}
and note the following identities
\begin{equation}
\int_0^{x_-}\frac{dx}{\sqrt{x^4-Ax^2+1}}\,=\frac{1}{x_+}\, K(x_-^4), \qquad
\int_0^{x_-}\frac{x^2\, dx}{\sqrt{x^4-Ax^2+1}}\,={x_+}\, [K(x_-^4)-E(x_-^4)], 
\end{equation}
where $A>0$, $x_-$ is the smallest positive root of the radicand (which has, per assumption, four real roots), and $x_-x_+=1$. Taking $v$ out of the radicand and introducing a new variable $z=u\sqrt{|J|/v}$ we obtain expressions of the above form with the result:
\begin{equation}
\De(J,v)=\frac{J+1}{\sqrt{v|J|}}\,\frac{1}{z_+}\, K(z_-^4)-\frac{\sgn(J) \sqrt{v}}{\sqrt{|J|}}\,z_+\left[K(z_-^4)-E(z_-^4)\right],   
\end{equation}
where $z_\pm$ are (positive) roots of
\begin{equation}
z_\pm^2=\frac{1}{2}\,\left[A\pm\sqrt{A^2-4}\right], \qquad \text{with }A=\frac{J+2}{v} \sgn(J)>0. 
\end{equation}
The total scattering angle is just $\th=\pi-2\De$. The scattering angles have an interesting expansion for large values of the impact parameter $b=(J+1)/v$: 
\begin{align*}
\De(b,v)& \approx  \frac{\pi}{2} + \frac{\pi (2 + v^2)}{8 v^2 b^2} - \frac{\pi}{
 2 v b^3} + O(b^{-4}), \qquad b\ra+\infty\\
-\De(b,v)& \approx \frac{\pi}{2} + \frac{\pi (2 + v^2)}{8 v^2 b^2} + \frac{\pi }{
 2 v b^3}+ O(b^{-4}), \qquad b\ra-\infty, 
\end{align*}
where the negative values of $b$ correspond to rays coming from below the $\hat x$ axis in the left part of Fig. \ref{geod}. The symmetry of the first two terms, as well as an overall focusing character of the vortex scattering\footnote{An angle $\De>\pi/2$ for rays coming from above $\hat x$ axis corresponds to a focusing behavior of the scattering.}, drew attention of authors before \cite{F_V_vortex}. On the ``positive absorption edge'' ($J=0$) we have $\De\ra+\infty$, while $\De\ra-\infty$ on the ``negative absorption edge'' ($J=-2v-2$). Note also, that due to $v^2<1$ there holds 
\begin{equation}
w_-=\frac{J+2-\sqrt{(J+2)^2-4v^2}}{2J}<\frac{1}{J}=w_{caus},
\end{equation}
and therefore geodesics with $J>0$ coming from $\infty$, which potentially could enter the region with $\dot t<0$, get reflected before reaching that region.


\section{Mathematical structure of wave-propagation problem}
In this chapter we consider the propagation of (density) waves in a fluid exhibiting a line-vortex flow following the acoustic spacetime approach \cite{Barcelo,\StoneWaveEq}. In this approach one notices that, as long as the background flow is irrotational, the perturbation $(\de \rho,\de \vec v)$ is parameterized by a single potential potential $\psi$. This potential fulfills a single partial differential equation having the form of the d'Alembert equation \eqref{dAlembert} in the acoustic spacetime, the metric of which in our case is given by \eqref{s_spacetime}.  While the most straightforward approach would consist of writing and separating this equation in order to determine the available ``sound modes'', we first present the mathematical structure behind the problem. On the one hand this provides us with connections to other problems of theoretical physics, and on the other helps to avoid missing important caveats which, in our opinion, have not received due attention in the literature.

\subsection{Geometrical preliminaries: constant $t$ foliation}
The acoustic spacetime has the line element
\begin{equation}
ds^2=-f\, dt^2-2\, dt\, d\vp+r^2\, d\vp^2 +dr^2+dz^2
\end{equation}
with $f=1-r^{-2}$. We have: $g\eqdef $det\,$(g_{ab})=-r^2$, and (displaying $(t,\vp)$ components only)
\begin{equation}
g_{ab}=\begin{pmatrix} -f & -1\\ -1 & r^2 \end{pmatrix}
,\qquad g^{ab}=-\frac{1}{r^2}\, \begin{pmatrix} r^2 & 1\\ 1 & -f \end{pmatrix}
\end{equation}
We may distinguish $t=const$ surfaces $\mathcal C_t$, which foliate the acoustic spacetime. The unit normal vector $N^a$ orthogonal to these surfaces, $N_a=-\d_a(t)$ is always timelike. The Killing vector of $t$-translations $T^a=(\d_t)^a$ may be orthogonally decomposed as
\begin{equation}
T^a=\a N^a+\be^a,
\end{equation}
with
\begin{equation}
\a=\frac{1}{\sqrt{-g^{tt}}}=1, \qquad  \be^a=(0,-\tfrac{1}{r^2},0,0), \qquad \be_a=(\tfrac{1}{r^{2}},-1,0,0),
\end{equation}
 finally
\begin{equation}
N^a=(1,\tfrac{1}{r^2},0,0).
\end{equation}
The vector fields: $e_T^a=N^a$ and $\axun^a=-r\be_a=(0,\tfrac{1}{r},0,0)$ form an orthonormal tetrad.

For every covariantly conserved current $J^a$ by Gauss theorem we have that
\begin{equation}
Q=\int_{\CC_t} dS^a\, J_a
\end{equation}
is independent of the foliation-parameter $t$, provided the fluxes through boundaries of $\cauchy$ are either absent or integrate to zero. The volume element of $\cauchy$ is $dS^a=\sqrt{h}\, d^3y\, N^a$, which for the special case of the vortex spacetime is just $d^3\v x\, \de^a_t$, because the internal geometry of $\cauchy$ is flat: 
\begin{equation*}
ds^2|_\cauchy=h_{ab}dx^adx^b=r^2d\vp^2+dr^2+dz^2.
\end{equation*}
Note, that in our case the determinants fulfill $-g=h=r^2$.

\subsection{Sound propagation as a linear dynamical system}\label{lin_dyn_sys}
\subsubsection{Function spaces}
In this section we provide some general arguments intended to fix the mathematical context of the subsequent more concrete investigations.  Let us consider generally the space of ``initial data'' of the evolution of the scalar potential $\psi(t,\vec x)$  in time $t$. Because the equation governing the evolution (the d'Alembert equation) is of second order in $\d_t$, the space of initial data on $\cauchy$ will necessarily contain the information about the potential $\psi$ and about its ``time-derivative'' $\PP=N^a\d_a\psi$. The standard structure of a linear dynamical system associated with this problem will be sketched below. Firstly we introduce the space of \emph{real} solutions of the wave-equation, $\mathcal S_\RR$. Because of the uniqueness of the time evolution, this space is equivalent to the space of initial data, i.e. to the space of  elements of the form $\Phi=(\psi,\PP)$ on any of the Cauchy surfaces $\cauchy$. In what follows we will make no distinction in notation between the space of solutions and the space of initial data (phase space). We equip the structure with the Poisson bracket 
\begin{equation}\label{symplectic form}
\s[\Phi_1,\Phi_2]=\int_{\CC_t} dS^a\, j_a[\Phi_1,\Phi_2]=\int d^3x\, (\psi_1\PP_2-\PP_1\psi_2),
\end{equation}
where the rightmost form is specific to the vortex acoustic spacetime, and $j_a$ stands for the covariantly conserved current
\begin{equation}
j_a[\Phi_1,\Phi_2]= \Phi_1 \d_a \Phi_2 - \d_a \Phi_1  \Phi_2.
\end{equation}
We also consider the space of complex solutions, or complex initial data, $\mathcal S_\CCC$, equipped with the same Poisson bracket (but with the first argument complex-conjugated). We mention, that the dynamics of the system will be governed by a Hamilotnian of equation \eqref{EOM}.

\subsubsection{Relation to quantum fields in curved spacetimes}

The above structures are sufficient from the point of view of classical field theory. At this point it is worthwhile to consider also the deeper problem of quantum sound fields propagating on a given flow of fluid, such as the field of phonons in a BEC. There are reasons to believe that quantized density fluctuations in fluids with phase coherence do exist and behave as a quantum field (see, however, \cite{Feynman_helium}). Indeed, this type of reasoning provides the basis for sonic analogs of Hawking-, Unruh- and other phenomena traditionally restricted for quantum fields in flat and curved spacetimes \cite{Visser_review}. 

 Quantization of fields usually consists of two steps: the relatively easy one\footnote{This step is relatively easy for free fields, which is the case we consider here.} concerned with the formulation of the algebra of field observables, and the more difficult one concerned with a construction of a well-behaved representation of this algebra on a Hilbert space (such as the Fock space). The construction of a ground-state representation for fields in stationary axisymmetric spacetimes without ergoregions was given by \pers{Kay} \cite{Kay_1p}. Technically what is needed, is a Schr\"odinger-type description of the evolution of the ``positive-frequency'' solutions of the field equation.  In order to achieve this one forms a complex \emph{Hilbert space} of (one-component) functions $\KK$, usually known under the name of one-particle Hilbert space. The elements of $\KK$ are meant to correspond to the  ``positive-frequency'' solutions in $\SS_\CCC$, with different choices of ``positive frequencies'' leading to quantized-field representations related by a Bogoliubov transformation. The time-evolution should be represented in $\KK$ as a unitary operator $U:\KK\ra\KK$.

In simple cases, such as the propagation of fields in Minkowski spacetime, the way to project elements of $\SS_\CCC$ into $\KK$ is provided by an operator $j:\mathcal S_\CCC\ra \KK$. For a solution ``of frequency $\w$'' the action of $j$  would have the form
\begin{equation}
j(\Phi)=\tfrac{1}{\sqrt{2\w}} (\w \psi - i \PP), 
\end{equation}
where $\PP=\d_t\psi$ in that case. In our case, however, the frequency corresponding to an element of the initial data space $\SS_\RR$ is not yet defined. The standard definition of the frequency is that it is the eigenvalue of the Hamilton-operator, $H$, which is $t$-independent for the stationary spacetimes. What is needed for the definition of the projection operator $j$, and thus also for the construction of the ground-state representation, is an operator $h$ such that $h\psi=\w \psi$, $h \PP=\w \PP$ for an eigenvector of the Hamiltonian $H$ of the form $(\psi,\PP)=\Phi_\omega$. Then, the operator $j$ will fulfill 
\begin{equation}
j[\Phi(t)]=e^{iht}j[\Phi(0)], \qquad \text{i.e. } U(t)=e^{iht}.
\end{equation}
 In the simplest case of a scalar field on the Minkowski space, the choice is $h=\sqrt{-\nabla^2}$, but we do not know if an appropriate choice can be made in the case of stationary spacetimes with ergoregions. We note, that the investigation of quantum sound fields can be carried further using methods developed by \pers{Kay} \cite{Kay_1p}, although we stress, that these methods will not be available if an ergoregion is present. In this latter case no ground-state representation of the above type is known, and only more abstract solutions of the problem are available (\cite{SS,Fulling_book}), the physical interpretation of which has yet to be explored.

\subsubsection{Time evolution}

Returning to the classical fields, and denoting by $\mathcal T(t)\Phi_0$ the time-evolution of a phase-space vector $\Phi_0\in\SS_\CCC$ we  derive the form of the Hamiltonian
\begin{equation}\label{EOM}
i\left.\frac{\d }{\d t}\,\Bigl[ \TT(t)\Phi\Bigr]\right|_{t=0}=H\Phi_0=-\s_2 A\, \Phi_0, \qquad \text{with } H=\begin{pmatrix}
i\be\d & i \\
i\nabla^2 & i\be \d
\end{pmatrix}, 
\end{equation}
where for technical reason we have distinguished the operator 
\begin{equation}
A=\begin{pmatrix}
-\nabla^2 & -\be^i\d_i\\
\be^i\d_i & 1
\end{pmatrix},
\end{equation}
while $\s_2$ stands for the Pauli matrix. Anticipating the separation of the angular dependence via $e^{im\vp}$ (with $m\in\ZZ$), leading to the \emph{partial-wave-expansion}, we give the form of $A$ for partial waves:
 \begin{equation}
A=\begin{pmatrix}
-\nabla^2 & \frac{im}{r^2}\, \\
-\frac{im}{r^2}\,  & 1
\end{pmatrix}
\end{equation}
where $\nabla^2$ is the Laplace operator (of a flat 3-dimensional space).

We note the connection to a technically very similar problem of complex scalar fields in electrostatic potentials, as investigated in \cite{SS}. Setting the (external) vector potential to zero ($\overrightarrow A=0$) there, and denoting the mass of the field by $\mu$, we obtain
\begin{equation} 
A=\begin{pmatrix}
-\nabla^2 +\mu^2& iV\\
-i V & 1
\end{pmatrix}\! ,
\end{equation} 
where $V$ is the scalar potential. The correspondence between \cite{SS} and our situation is therefore: $V\equiv i\be\d$, which for partial waves reduces to $V\equiv m/r^2$. Thus, from the hamiltonian point of view, the behavior of (complex) sound potential in a vortex-background is equivalent to the behavior of a charged scalar field in strongly attractive/repulsive (depending on $m$) potentials. Such problems have been investigated in the literature before, see e.g. \cite{\Fulling1} and the appendix B of \cite{Fulling_book}). 

For the linear space of Cauchy data one may introduce a natural, positive-definite, product
\begin{equation}
\langle \Phi,\Phi\rangle=\int_\cauchy d^3y\, \sqrt{h}\, \Phi^\dagger \Phi=\int d^3\v x\ \Phi^\dagger \Phi,
\end{equation}
and also the product resulting from the Poisson bracket between $\Phi^\dagger$ and $\Phi$, which (conforming to the literature) we shall call ``the norm of $\Phi$'',
\begin{equation}\label{Kproduct}
\|\Phi\|^2=(\Phi,\Phi)_\HH\eqdef-\langle \Phi, \s_2 \Phi\rangle=i \int_\cauchy d^3y\,\sqrt{h}\, (\ba \psi \PP -\ba \PP \psi)=i \s [\ba \Phi,\Phi]=2 \int d^3 \vec x\, (\w-\tfrac{m}{r^2})|\psi_\w|^2,
\end{equation}
(the last equity for $\psi$ of the form $e^{-i\w t+im\vp}\psi_\w(r)$). The space $\SS_\CC$ equipped with the product $\langle .,.\rangle$ is a Hilbert space, while the space $\HH$ equipped with $(.,.)_\HH$ is a Krein (indefinite-product) space. This product corresponds exactly to equation (34) of \cite{SS} (with $\PP=(\d_t-iA_0)\psi$ there), and to (2.15) of \cite{Fulling_book}. The merit of introducing these additional function spaces lies in the special properties of the operators $H$ and $A$.

\subsubsection{Properties of the Hamiltonian}
The Hamilton operator, $H=-\s_2A$, which is explicitly given by
\begin{equation}\label{hamiltonian}
H=\begin{pmatrix}
i\be\d & i \\
i\nabla^2 & i\be \d
\end{pmatrix}, \qquad \text{for partial waves:}\quad
H=\begin{pmatrix}
\frac{m}{r^2}\,  & i \\
i\nabla^2 & \frac{m}{r^2}\, 
\end{pmatrix}
\end{equation}
is a symmetric operator on the space $\HH$, i.e. $(\Phi,H\Psi)_\HH=(H\Phi,\Psi)_\HH$. Additionally, in some cases, the operator $A$ is a positive operator w.r.t. the product $\langle .,.\rangle$ (that is $\langle \Phi,A\Phi\rangle\geqslant 0$ for all $\Phi\in\SS_\CCC$). In these cases one may proceed as \pers{Kay}, and define the Hilbert (positive-product) space $\mathcal A$ with the product $
\langle ., .\rangle_{\mathcal A}$
\begin{equation}
\langle \Psi, \Phi\rangle_{\mathcal A}=\langle \Psi, A \Phi\rangle, 
\end{equation}
on which the Hamiltonian $H=-\s_2A$ could be shown to be selfadjoint. In such cases, therefore, $H$ has the standard properties of selfadjoint operators in Hilbert spaces, such as: the realty of eigenvalues $\w$, the orthogonality of the eigenvectors, and the completeness of the family of eigenvectors. All these properties, therefore, hold for sound modes propagating on a vortex background provided that the boundary conditions at the innermost part of the vortex lead to a self-adjoint $H$.

In the cases where $A$ is not positive definite, one is left with the Krein space $\HH$, and with weaker results for $H$ (see appendix \ref{J-adjoint}), in particular: eigenvalues are not necessarily real (but do come in pairs of complex-conjugate numbers) and a broader class of vectors is necessary for completeness. As we shall see below, the distinction between cases - for stationary axisymmetric spacetimes - corresponds exactly to the distinction between spacetimes with/without ergoregions.

\subsection{Energy and angular momentum}
Let $\xi_T^a=(\d_t)^a$ denote the Killing vector corresponding to time translation. The current associated with the notion of energy has the form
\begin{equation}
J^a_T[\psi]=?T^a_b?[\psi] \xi^b_T
\end{equation}
where $T^{ab}[\psi]$ stands for the energy-momentum tensor of the solution $\psi(t,\vec x)$:
\begin{equation}
T_{ab}[\psi]=\d_{(a}\ba \psi\d_{b)}\psi-\frac{1}{2}\,g_{ab}\, g^{cd}\, \d_c\ba \psi\d_d\psi, 
\end{equation}
and the index-brackets denote symmetrization, $A_{(ab)}=\tfrac{1}{2}(A_{ab}+A_{ba})$. The energy is obtained by integrating $J^a_T$ over any Cauchy surface $\cauchy$,
\begin{equation}
E[\psi]=\int_{\CC_t} dS_a J_T^a[\psi].
\end{equation}

 Note, that formally (if relevant boundary terms vanish) 
\begin{equation}\label{EnergyNormRelation}
E[\Phi_\w]=\tfrac{1}{2}\langle \Phi_\w,A\Phi_\w\rangle=-\tfrac{1}{2}\langle\Phi_\w,\s_2 H\Phi_\w\rangle=\tfrac{1}{2}(\Phi_{\w},H \Phi_{\w})_\HH=\jd {\w}||\Phi_\w||^2.
\end{equation}
For scalar fields in Minkowski space the norm $||\Phi_\w||^2$ corresponding to positive frequency ($\w>0$) solutions is always positive; here this does not need to be the case (because positivity of $E[\Phi_\w]$ is not guaranteed).

We may also regard energy as a functional of the solutions of the wave equation $\psi(t,\vec x)$, 
\begin{multline}\label{energy}
E[\psi]=\int d^3\v x\, T_{ab}N^aT^b =\tfrac{1}{2}\,\int d^3\v x \left\{|\d_t\psi|^2+\tfrac{1}{r^2}\left(1-\tfrac{1}{r^2}\, \right) |\d_\vp\psi|^2+|\d_r\psi|^2+|\d_z\psi|^2 \right\}=\\=\tfrac{1}{2}\,\int d^3\v x \left\{|\nabla \psi|^2+|\PP|^2+\tfrac{1}{r^2}[\ba \psi \d_\vp \PP-\ba \PP \d_\vp \psi]\right\}=\tfrac{1}{2}\langle \Phi, A\Phi\rangle, 
\end{multline}
where the integration by parts, $-\int d^3\v x\, \ba \psi (\nabla^2\psi)=\int d^3\v x |\nabla \psi|^2$, was performed. (All boundary conditions employed in what follows are such that the - a priori present - boundary terms do vanish.) The energy functional is evidently positive when $r$ is bounded from below by a $r_0>1$, i.e. in the absence of ergoregions. 

The angular momentum associated with a field configuration $\psi$, $J[\psi]$, will be found as the Cauchy-surface integral of the current associated with the Killing vector field corresponding to rotations, $\xi_\Phi^a=(\d_\vp)^a$, 
\begin{equation}
J[\psi]=\int d^3\v x\, T_{ab}\, N^a \xi^b_{\Phi}=\int d^3\v x\ \text{Re}(\ba \PP \d_\vp \psi) =m\int d^3\v x\, (\w -\tfrac{m}{r^2})|\psi|^2,   
\end{equation}
with the last equity holding for solutions of the form $e^{-i\w t+im\vp}\psi(r)$, for which we also find
\begin{equation}\label{J_Norm}
J[\psi]=\jd m\,||\Psi||^2. 
\end{equation}


\section{Partial wave decomposition}

The scalar d'Alembert equation reads
\begin{align}
\Box \psi&=\frac{1}{\sqrt{-g}}\d_i\left(g^{ij}\sqrt{-g}\d_j\psi\right)=\notag \\ &=\frac{1}{r}\,\d_r(r)\, \d_r\psi+g^{ij}\d_i\d_j\psi=\frac{1}{r}\,\d_r\left(r\d_r \psi\right)+\left[-\d^2_t-\frac{2}{r^2}\d_{t\vp}+\left(\frac{1}{r^2}-\frac{1}{r^4}\right)\, \,\d^2_{\vp} +\d^2_z\right]\psi=0\label{dAlembert}
\end{align}
and is equivalent to a Schr\"odinger-type equation
\begin{equation}\label{hamilton}
i\d_t \Phi = H\Phi, \qquad \text{ with }\Phi_\w=(\psi,\PP)\in \HH.
\end{equation}
Thus, the problem of finding ``modes'' either simplifies solely to the eigenvalue problem for the Hamilton operator
\begin{equation}\label{eigenproblem}
H\Phi_\w=\w \Phi_\w, 
\end{equation}
or (in case $H$ is not a normal operator) leads also to equations for Jordan associated  eigenvectors
\begin{equation}
H\Phi^a_\w=\w \Phi^a_\w + \Phi^{a-1}_\w,
\end{equation}
with a (potentially infinite) chain numbered by the index $a\in\mathbb N$ and $\Phi^0_\w=\Phi_\w$. The Jordan eigenvectors evolve in time via
\begin{equation}
\Phi^a_\w(t)=e^{-i\w t}\left[\Phi^a(0)-it\Phi^{a-1}(0)\right],
\end{equation}
fulfill the equation \eqref{hamilton} and therefore are associated with solutions of the \emph{homogeneous} wave equation $\Box \psi=0$. If we had a finite-dimensional case at hand, the associated vectors would be necessary to obtain a complete set of vectors in the linear Hilbert space with the product $\langle.,.\rangle$. We presume this also to be the case in the present context, should Jordan vectors exist for the given $H$ with chosen boundary conditions.

Proceeding with our special problem, we separate the angular- and the $z-$dependence, as is usually done, by
\begin{equation}
 \Phi(t,\vec x)=e^{i(m \vp + p z)}\Psi(t,r), \qquad m\in\mathbb Z,\ p\in\mathbb R.
\end{equation}
The domains of $m$ and $p$ are justified by single-valuedness of $\Phi$ and self-adjointness of $-i\d_z$, respectively. The problem is stationary, and for this reason we separate the time-dependence by
\begin{equation*}
\Psi(t,r)=e^{-i\w t}\Psi_\w(r),
\end{equation*}
where, however, $\w$'s are not a priori restricted to be real numbers.  Let us consider the eigenvalue problem for $H$, eq. \eqref{eigenproblem}, in a sector of fixed $(m,p)$. The $\HH$-product, eq. \eqref{Kproduct}, of two eigenvectors of $H$ assumes the form
\begin{equation}\label{product}
(\Phi_{\w_1},\Phi_{\w_2})_\HH=\int d^3 x\, (\ba{\w_1}+\w_2-\tfrac{2m}{r^2})\ba{\psi_{\w_1}}\psi_{\w_2},
\end{equation}
where $\Phi_\w=\bigl(\psi_\w,-i(\w-\tfrac{m}{r^2})\psi_\w\bigr)$.

The eigenvalue problem reduces to solving the radial equation for $\psi\equiv \psi_\w$, which is the main equation of this paper:
\begin{equation}\label{main_eq}\boxed{
\psi''+\frac{1}{r}\,\psi'+\left[k^2-\frac{M^2}{r^2}\,+\frac{m^2}{r^4}\,  \right]\psi=0, }
\end{equation}
with\footnote{The equation(-s) leading to the Jordan associated eigenvectors (elements of the Jordan chain) is an inhomogeneous one: $$\left(\d^2_r+\frac{1}{r}\d_r\right)\psi_a+\left[k^2-\frac{M^2}{r^2}\,+\frac{m^2}{r^4}\,  \right]\psi_a=2\left(\frac{m}{r^2}\,-\w \right)\psi_\w,$$
where $\psi_\w$ is either the solution of the homogeneous equation, or the solution at the previous position in the Jordan chain, as the problem has the form $(H-\w)\psi_{a+1}=\psi_{a}$.
}
\begin{equation*}
k^2=\w^2-p^2, \qquad M^2=m(m+2\w).
\end{equation*}
Neither $M^2$ nor $k^2$ need to be positive; negative values of these parameters will correspond to  physically distinguished situations. From now on we will use the notation
\begin{equation*}
D^2_r=\frac{d^2}{d r^2}\,+\frac{1}{r}\, \frac{d}{d r}\,  
\end{equation*}
for the (often appearing) radial part of the two-dimensional Laplace operator. The frequency, $\w$, as an eigenvalue of symmetric operator, $H$, on a Krein space does not need to be real. The radial equation has the symmetry $(m,\w)\ra(-m,-\w)$, so that finding all positive frequency solutions would be enough. We note the singularity at $r=0$, which in the usual case (i.e. if we were dealing with the usual Hilbert space of square-integrable functions on $\RR^3$) leads to well-known ambiguities associated with selfadjoint extensions of the Hamilton operator\cite{\citCase}.

If not for the scalar product, the problem would be of the Schr\"odinger form with
\begin{equation}\label{schr}
-(D^2_r\psi)+\underbrace{\left[\frac{M^2}{r^2}-\frac{m^2}{r^4}\right]}_{V(r)}\psi=k^2\psi,
\end{equation}
with the potential $V(r)$ diverging to $-\infty$ as $r\ra0$.

Momenta along the vortex, $p$, are not necessarily restricted by $p^2\leqslant \w^2$. Such a restriction in usual scattering problems comes from the required behavior the innermost parts of the ``space'', for example from a boundary condition at a fixed $r$, or from a regularity requirement at $r=0$. We will investigate this problem in section \ref{section_boundary_condition}. 
 
For comparison with ray acoustics, we note the relations
\begin{equation}
\dot t=E-\frac{J}{r^2}, \qquad \dot \vp = \frac{E+J(1-\tfrac{1}{r^2})}{r^2},
\end{equation}
and
\begin{equation*}
\dot r^2=(E^2-p^2)-\frac{J(J+2E)}{r^2}+\frac{J^2}{r^4},
\end{equation*}
so that for apt analogy with \eqref{schr} we should make identifications $m\leftrightarrow J$, $\w\leftrightarrow E$, and $k^2=E^2-p^2$. We also note the correspondence between $i\PP \psi=(\w-\tfrac{m}{r^2})\psi$ and $\dot t$ of the acoustic rays.

\section{Boundary condition at/near the vortex core}\label{section_boundary_condition}
The Hamiltonian is defined by its formal expression, \eqref{hamiltonian}, together with appropriate boundary conditions. We recall, that in the case of operators in Hilbert spaces choosing a ``wrong'' boundary conditions can make a symmetric operator (fulfilling $\langle\psi,H\chi\rangle=\langle H\psi,\chi\rangle$) non-selfadjoint, opening-up the possibility of complex frequencies $\w$. For a one-dimensional differential operator posing a boundary condition at a regular point of the corresponding ODE is relatively unproblematic. Moreover, the case of  boundary conditions supposed to be posed at singular points of the ODE there is also a well established theory due to \pers{Weyl} \cite{Weyl}. In the vortex case, assuming the whole space $r\in[0,\infty)$ is available, we have the problem of posing the boundary condition at $r=0$. This point is a singular point of the radial equation, and the Hamiltonian is not given as a symmetric operator on a Hilbert space.

Let us recall an important example. For the Schr\"odinger problem (in Hilbert space) with the potential \mbox{$V=-\tfrac{m^2}{r^4}$} \pers{Case} proved \cite{\citCase}, that for small $r$ the eigenfunctions \emph{necessarily} have the asymptotic form \mbox{$\psi\approx r\cos\left(\tfrac{|m|}{r}+B\right)$} with \emph{the same value of the phase $B$} for all $\w$. Posing a boundary condition at $r=0$ is equivalent to a choice of $B$, which then becomes the essential property of the operator. This inhomogeneous condition replaces the more commonly employed homogeneous boundary conditions (such as the Dirichlet/Neumann conditions)  at $r=0$, and it is necessary for the  \emph{mutual orthogonality} w.r.t. $\langle.,.\rangle$ of eigenfunctions to different eigenvalues (a ``must'' for symmetric operators on a Hilbert and Krein spaces)\footnote{ Misprints in eq. (26) of \cite{\citCase} do not influence this result; the power of $r$ multiplying the cosine function is special to \pers{Case}'s form of the problem.}. A reasoning, similar to what has been done in \cite{\citCase}, can be repeated in our case. 

Close to $r=0$ the radial equation assumes the form, 
which it always has for $k=0$ (i.e. in the case of waves the $\hat z$ momentum equal to the frequency, $p=\w\in\RR$). Introducing $u=1/r$ we find,
\begin{equation}
u^2\psi''+ u\psi'+(m^2u^2-M^2)\psi=0, 
\end{equation}
and thus
\begin{equation}
\psi=\Bessel{M}{|m|/r},\qquad \text{for }r\ll 1\text{ or for }k=0,
\end{equation}
where $\Bessel{M}{x}$ stands for an arbitrary regular Bessel function ($J,Y$ or $H^{1,2}$) of order $M$. We recall that even for $\w\in\RR$ the parame\-ter $M$ can be real or imaginary (the latter case occurs for \mbox{$m\in(-2\w,0)$}, which is the wave analog of ray-capture\footnote{Restoring the arbitrary value of $E$, the acoustic rays with $E^2\approx p^2$ would reach the core for $J\in[-2E,0]$.}). 

The asymptotic expansions of the radial function at $r\ll 1$ for the distinguished Bessel functions read
\begin{align}
J_M(\tfrac{|m|}{r})&\approx \ \ \cos\left(\tfrac{\pi}{4}+\tfrac{M\pi}{2}-\tfrac{|m|}{r}\right)\cdot \sqrt{\tfrac{2r}{\pi}},\\
Y_M(\tfrac{|m|}{r})&\approx -\! \sin\left(\tfrac{\pi}{4}+\tfrac{M\pi}{2}-\tfrac{|m|}{r}\right)\cdot \sqrt{\tfrac{2r}{\pi}}.
\end{align}
For any solution of the radial equation \eqref{main_eq}, $\psi$, the asymptotic expansion at small $r$ can be characterized by a phase $B$:
\begin{equation}
\psi\approx \cos B \cdot J_M(\tfrac{|m|}{r})-\sin B \cdot Y_M(\tfrac{|m|}{r})\approx \cos\left(\tfrac{\pi}{4}+\tfrac{M\pi}{2}-\tfrac{|m|}{r}-B\right)\cdot \sqrt{\tfrac{2r}{\pi}},
\end{equation}
or (shifting $B$ by $\tfrac{\pi}{4}+\tfrac{M\pi}{2}$):
\begin{equation}\label{as_expansion}
\psi\approx \cos(\tfrac{|m|}{r}+B)\cdot \sqrt{\tfrac{2r}{\pi}\, }\, . 
\end{equation}
We now repeat \pers{Case}'s reasoning in order to reveal the type of boundary condition at $r=0$ necessary in our case. 
Let $\Phi_1$ and $\Phi_2$ be two such solutions corresponding to the frequencies $\w_1$, $\w_2$, and characterized by phases $B_1$ and $B_2$ close to the core. Assume, that both correspond to \emph{the same} values of the separation constants ($m,p$). In order to simplify matters at $r=\infty$ we assume both of the functions to fulfill a Dirichlet boundary condition at some large $r=R$, and investigate them on the interval $r\in[0,R]$.  Multiplying the radial equations \eqref{main_eq} fulfilled by each function by the other function, subtracting them, and integrating the result from $0$ to $R$ we obtain
\begin{equation}
\left. r\left[\ba{ \psi_1} \d_r \psi_2 -\ba{\d_r \psi_1}\psi_2\right]\right|_0^R=(\w_2-\ba{\w_1})(\Phi_1,\Phi_2)_\HH.
\end{equation}
(More precisely, there should stand only the radial part of the $\HH$-product on the r.h.s..) Because eigenfunctions $\HH$-selfadjoint operator to different eigenvalues ($\w_1,\w_2$) must be $\HH$-orthogonal (see appendix \ref{J-adjoint}), using the asymptotic expansion at $r=0$, eq. \eqref{as_expansion}, we arrive at
\begin{equation}
|m|\sin(B_1-B_2)=(\w_2-\ba{\w_1})(\Phi_1,\Phi_2)_\HH\equiv0,
\end{equation}
where on the l.h.s. a (regular) limit $r\ra 0$ has been performed. Thus, for each partial-wave problem (fixed ($m,p$)), the phases $B$ must be independent of the frequency $\w$. This is so unless we are dealing with a pair corresponding to mutually conjugate, complex frequencies, which - however - might appear in the problem. The boundary condition, as far as it can be fixed by the orthogonality-argument above, amounts to setting the phase $B(m,p)$ (a function of $m$ and $p$) to be independent of $\w$. In real physical situations the values of the phases $B(m,p)$ will be related to the physics of the innermost parts of the vortex, and will depend on the model of the core used, e.g. on the type of the fluid in motion. In the subsequent chapters we will \emph{illustrate} the fact, that the choice of these phases does influence essential physical characteristics of sound-propagation on the vortex. We will do so, by taking the $B(m,p)$ to be independent of $(m,p)$, essentially choosing either the Bessel- $J$ or $Y$ asymptotics for all $m$ and $p$. We do not expect this type of choice to be related to any particularly distinguished physical situation.

\section{Solutions to the radial equation}
\subsection{Structural results}
\subsubsection{General considerations}\label{general_con}
A general-enough consideration of the  radial equation, eq. \eqref{main_eq}, assumes only $m\in\mathbb Z$ and $p\in \mathbb R$. It is important to restrict the number of parameters to an essential set. Recalling, that a priori $\w\in\mathbb C$, and thus $k^2=\w^2-p^2$ is in general a complex number, we introduce a new variable $x^2=\left|\tfrac{k}{m}\right| r^2$, $x\in[0,\infty)$ and transform the radial equation \eqref{main_eq} into the form
\begin{equation}\label{mathieu_complex}
D^2_x\psi-\frac{M^2}{x^2}\,\psi+K^2\left[e^{i\, \arg(k^2)}+\frac{1}{x^4}\, \right]\psi=0, 
\end{equation}
where $M^2=m(m+2\w)$, and $|mk|=K^2$. We note, that by a simple transformation of variables the equation can be shown to be equivalent to the Mathieu equation, taken along various contours in the complex plane (see appendix \ref{mathieu_sec}). In the case of real frequencies the equation assumes the form(s) 
\begin{equation}\label{mathieu}
D^2_x\psi-\frac{M^2}{x^2}\,\psi+K^2\left[\pm 1+\frac{1}{x^4}\, \right]\psi=0, 
\end{equation}
where the $+$ sign corresponds to solutions having an interpretation of scattering states, occurring for $k^2=\w^2-p^2>0$, while the $-$ sign to ``bound states'', possibly occurring for $k^2<0$. Generally, unless $k^2$ is real positive, exactly one of two linearly independent solutions will be exponentially damped at $x\ra\infty$.

 Let us consider the scattering, $k^2>0$, case first. On the two extremes of the domain of $x$ ($0$ and $\infty$) we drop the fastest decaying  terms: $x^{-4}$ or $x^4=u^{-4}$, and obtain two Bessel equations with equal indices
\begin{align*}
D^2_x\psi-\left[\frac{M^2}{x^2}\,-K^2 \right]\psi &=0, \qquad x\gg 1,\\
D^2_u\psi-\left[\frac{M^2}{u^2}\,-K^2 \right]\psi &=0, \qquad u\gg 1,
\end{align*}
where $u=x^{-1}$.

\subsubsection{Bound states in the $(K^2,M^2)$ plane}\label{bound_section} 
Having established the general form of the radial equation we now study its distinguished solutions more closely. As remarked before, in the case of negative $k^2=\w^2-p^2$, the equation has the form
\begin{equation}
D^2_x\psi-\frac{M^2}{x^2}\psi+K^2\left[-1+\frac{1}{x^4}\, \right]\psi=0, 
\end{equation}
with $x\in(0,\infty)$. The solutions of this equation behave as modified Bessel functions ($K_{M}(K x)$ and $I_{M}(K x)$) at $x\ra\infty$. An important question arises as to whether for a given boundary condition at a finite (or zero) $x$ there exist values of $M$ and $K$, for which the solution would be ``bound-state-like'' with the asymptotics containing only the exponentially damped Bessel function. One might expect to be able to answer this question with the help of the specific investigations of the Mathieu equation (e.g. \cite{Meixner1,Meixner2,Abramowitz}). So far, however, we have not been able to locate the answer in the literature. For that reason we present here numerical results, from which important physical conclusions can be drawn. We shall use two types of boundary conditions:
\begin{enumerate}
\item[I)] The family of ``core''-boundary conditions provided by the orthogonality argument of section \ref{section_boundary_condition}. The values of the constant $B$ corresponding to asymptotic functions of the first ($J$) and second ($Y$) kind will be considered. We will only consider the (illustrative) cases, where the phase $B$ is independent of $m$ and $p$; otherwise the boundary condition would depend on the position in the $(K^2,M^2)$ plane.

\item[II)] Recalling that in the acoustic spacetime formalism the perturbation of the velocity field is just the gradient of $\psi$, we consider the family of Neumann boundary conditions posed at a finite $r=r_0$ (corresponding to a flow around, say, a central wire of radius $r_0$). This allows us to consider the effect of the presence/absence of an ergoregion, which is included in the spacetime if points with $r<1$ belong to it. \end{enumerate}

\subsubsection{Scaling transformations, and positions of bound states}
Both types of boundary conditions, the Neumann condition and the ``core'' condition \eqref{as_expansion}, can (for given values of regular parameters $(m,p,\w,r_0 \text{ or } B)$) be written purely in terms of the essential parameters/variables $(x,M,K)$. For the Neumann condition we require
\begin{equation}
\left. \frac{\d \psi}{\d x}\right|_{x=x_0}=0
\end{equation}
(with the $x_0$ computable from $r_0$, and from the other regular parameters), whereas for the ``core'' condition we require
\begin{equation}
\psi\approx \cos(\tfrac{K}{x}+B)\sqrt{x}.
\end{equation}
With this reformulation performed, we consider the radial (Mathieu) problem in the space of essential parameters. The results (such as positions of bound states in the $(K^2,M^2)$ plane) can always be recalculated to the space of regular parameters. An important ambiguity, however, occurs: because the number of essential parameters: $(x_0,K,M)$ for the Neumann condition, and $(B,K,M)$ for the ``core'' condition, is smaller than the number of regular parameters, $(r_0,m,p,\w)$ or $(B,m,p,\w)$, there exists a one-parameter family of transformations for the regular parameters, leaving the essential parameters invariant. Recalling the relations
\begin{equation}
x^2=\left|\tfrac{k}{m}\, \right| r^2, \qquad K^2=|mk|, \qquad M^2=m(m+2\w)
\end{equation}
we see immediately that essential parameters remain invariant if $(r_0,m_0,k_0,\w_0)$ is transformed into 
\begin{equation}
r_\la=\la r_1, \qquad m_\la=\la m_1, \qquad k_\la=\tfrac{1}{\la}\, k_1,\qquad \Om_\la=\tfrac{1}{\la^2}\,\Om_1, 
\end{equation}
where
\begin{equation}
\Om_\la=1+\tfrac{2\w_\la}{m_\la}. 
\end{equation}
The only restriction on $\la$ is, that $m_\la$ should also be an integer. The frequency transforms according to 
\begin{equation}
\w_\la=\tfrac{m_1}{2}(\tfrac{1}{\la}-\la)+\tfrac{\w_1}{\la}.
\end{equation}

\subsubsection{Positive/negative frequency bound states}

The $\la$-transformations allow to flip the sign of the frequency of a bound-state solution. Intro\-ducing an abbreviation $\mu_1=\tfrac{m_1}{2\w_1}\,$ we consider
\begin{equation}
\u \w_\la \eqdef \tfrac{\w_\la}{\w_1}=\mu_1(\tfrac{1}{\la}-\la)+\tfrac{1}{\la}. 
\end{equation}
An inspection of the above function shows, that unless $\mu_1\in[-1,0]$ (i.e. unless the $(\w_1,m_1)$ parameters fulfill $m_1\in[-2\w_1,0]$, which is equivalent to $M^2\leqslant 0$) the range of the function is the whole real set $\mathbb R$, and therefore the sign of the frequency will be flipped there, where $\u \w_\la<0$. The change of the sign of $\u \w_\la$ occurs for large $\la$ if $\mu_1>0$ and for small $\la$ if $\mu_1<-1$. Thus a solution corresponding to $\w_1>0,m_1>0$ can always be $\la$-transformed to produce a solution with $\w_\la<0,m_\la>0$. This is true in the ``existential sense'', i.e. it allows us to state that such solutions would exist in certain boundary-value problems. The $\la$-transformation for the Neumann boundary condition changes the radius $r_0$ (where the condition is posed), and therefore changes the problem considered. Solutions with the same sign of $\w$ and $m$ rotate in the positive direction (co-rotate with the vortex).

\subsubsection{ Negative norm states for positive frequencies}

For principal reasons in is important to investigate, whether positive frequency solution could in some cases correspond to negative ``norms'':
\begin{equation}\label{norm}
||\Phi_{\w}||^2=2 \int d^3 \vec x\, (\w-\tfrac{m}{r^2})|\psi_\w|^2,
\end{equation}
 With this regard let us stress, that for \underline{positive} $M^2$ the continuous nature of the positions of bound states in the $(K^2,M^2)$ plane allows us to investigate cases where $M^2=m^2$, with $m\in\mathbb Z$. Because of $M^2=m(m+2\w)$ we may regard $\w$ to be a linear function of $M^2$ indexed by the branch-numbering, discrete parameter $m$,
\begin{equation}
\w_m(M^2)=\tfrac{1}{2}\,\left(\tfrac{M^2}{m}-m \right). 
\end{equation}
Then, $\w_m(M^2)$ crosses zero at $M^2=m^2$, as we move along the bound-state-curve in $(K^2,M^2)$ plane. It is clear from \eqref{norm}, that in regions of negligible $\w$ and positive $m$ the ``norm'' of the corresponding solution will be negative. Therefore, states with $M^2$ slightly greater than the branch-defining $m^2$ will have positive frequency $\w_m(M^2)$ and negative norm $||\Phi_{\w_m(M^2)}||^2$. (Whether or not the norm of such a solution will turn positive as we continue along the bound-state-line in the $(K^2,M^2)$ is not clear.) 

On the other hand, the states with \u{negative} $M^2$ always correspond non-zero frequencies; the lowest possible frequency $\w_m(M^2$) for a given negative $M^2$ is $\w=\sqrt{-M^2}$ (this happens only in case $m=-\w$ is an integer). The question whether or not negative norm solutions for positive frequencies exist for $M^2<0$ does not appear to be answerable by general arguments of the above type.\\[2pt]

\subsubsection{\it Bound states and scaling transformations}

The $\la$-transformations allow us to tell, whether bound states are necessarily related to the presence of an ergoregion in the acoustic spacetime. We will consider the Neumann conditions, posed at $r_0$, with ergoregions disappearing from the spacetime once the condition is posed at $r_\la>1$. 

Let us take one solution related to a bound state, with the Neumann condition at $r_1$, and consider the $\la$-transformations. It is clear, that following the $\la$-trajectory, the new position of the boundary, $r_\la=\la r_1$ will - for sufficiently large $\la$ - be greater than $1$. Focusing on a fixed ``original problem'' with the parameters $(r_0,m_0,k_0,\w_0)$ we notice, that for very large $\la$'s the following limits arise: $m_\la\gg 1$, $k^2_\la\ll 1$ that is $p^2_\la\approx \w^2_\la$, with the frequency $\w_\la=-\tfrac{m_\la}{2}$ (and therefore $|\w_\la|\gg 1$). Such states, therefore, correspond to disturbances rotating in negative direction (against the vortex-rotation), with large angular quantum numbers, large frequencies and equally large momenta along the vortex.

\begin{figure}[htb]
\centering
\includegraphics[scale=0.55]{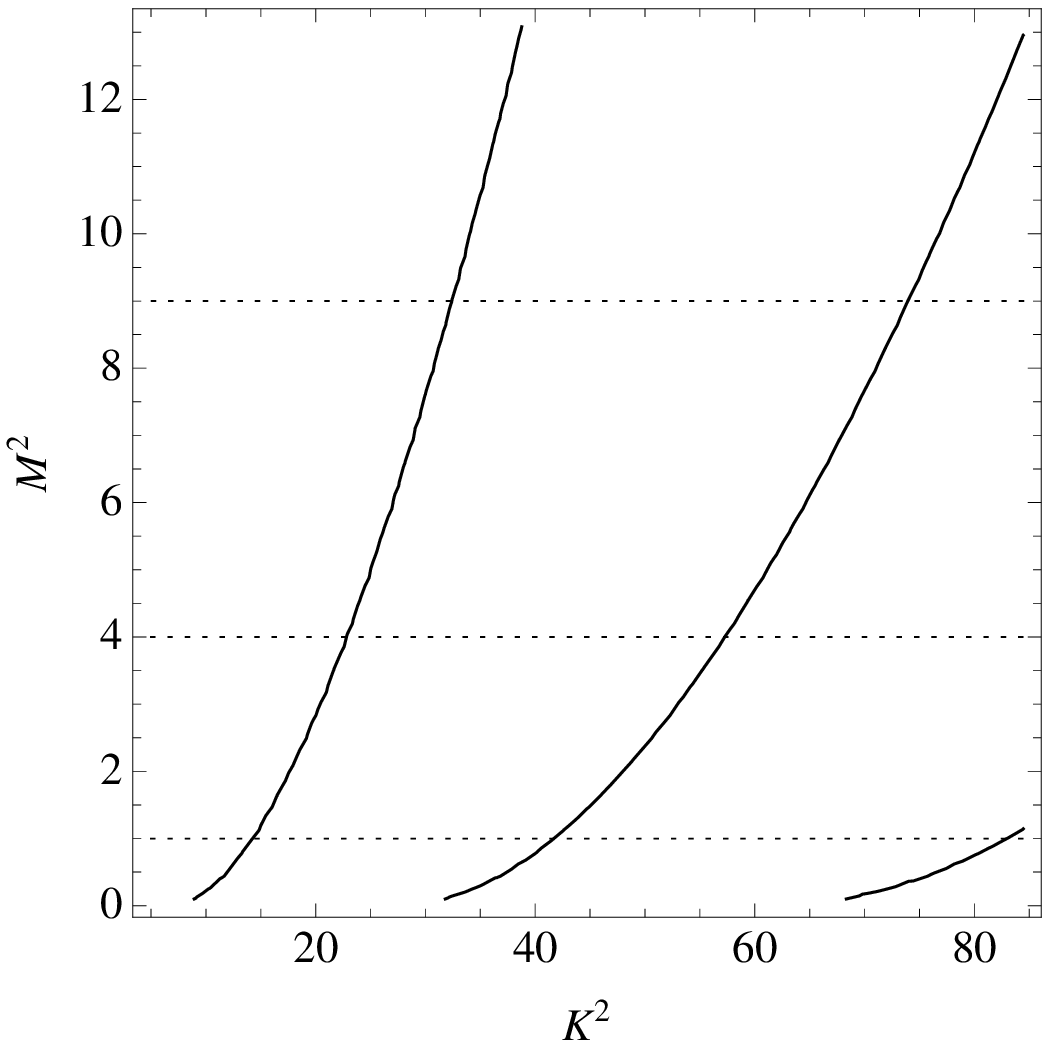}\qquad
\includegraphics[scale=0.55]{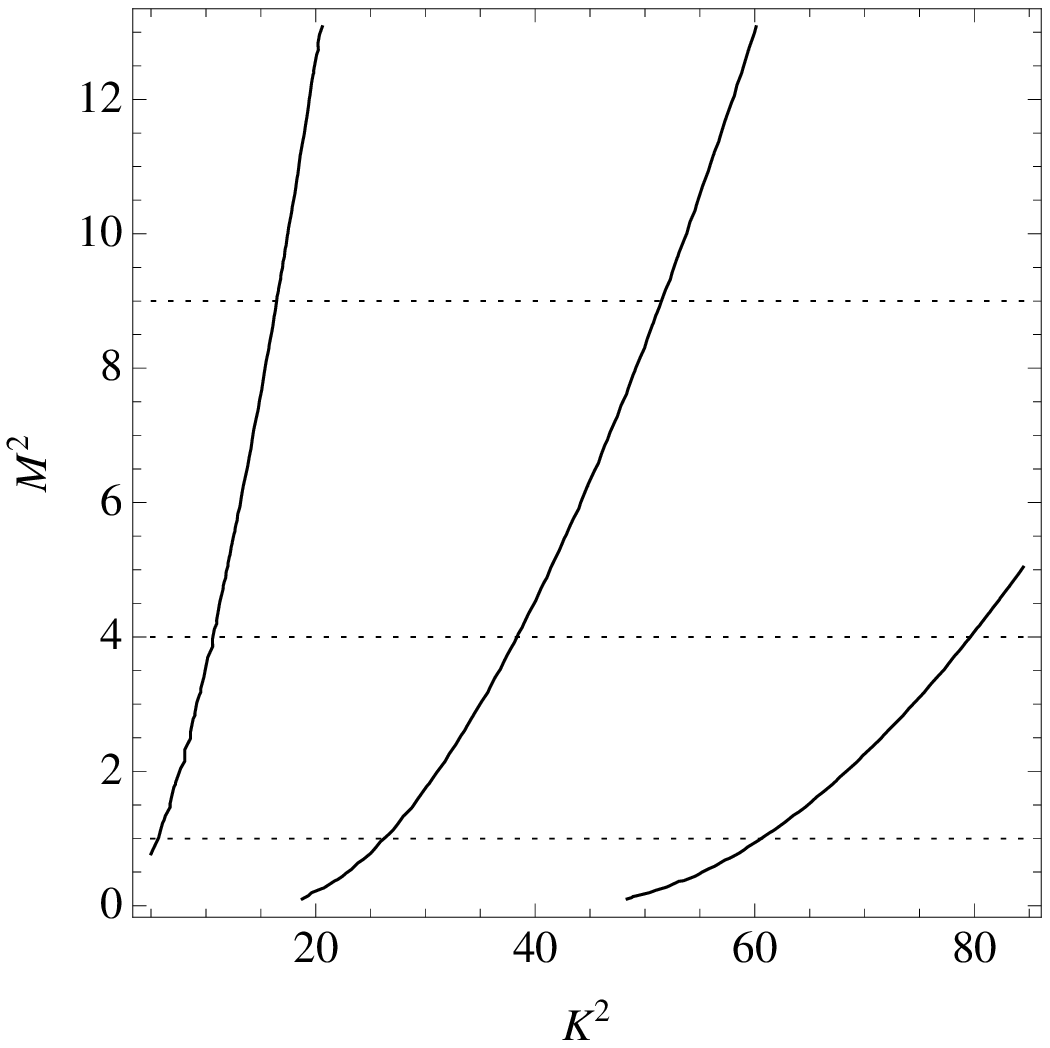}
\caption{Positions of the bound states in the $(K^2,M^2)$ plane for \emph{core boundary conditions}, together with distinguished lines $M\in\mathbb N$. [Left:] for the boundary condition $\psi\approx J_{M}(K/x)$, [Right:] for $\psi\approx Y_{M}(K/x)$. The lines are zeros of the Wronskian of the exact (numerical) solution with the Bessel function $K_{M}(Kx)$, which is the asymptotic form of the exponentially decaying solutions at $x=\infty$. We note, that none of the branches seems to enter/exist in the $M^2\leqslant 0$ quadrant, which suggests, that there are no bound states there (for the ``core'' boundary conditions).}
\label{fig_bs}
\end{figure}

\begin{figure}[h]
\centering
\includegraphics[scale=0.54]{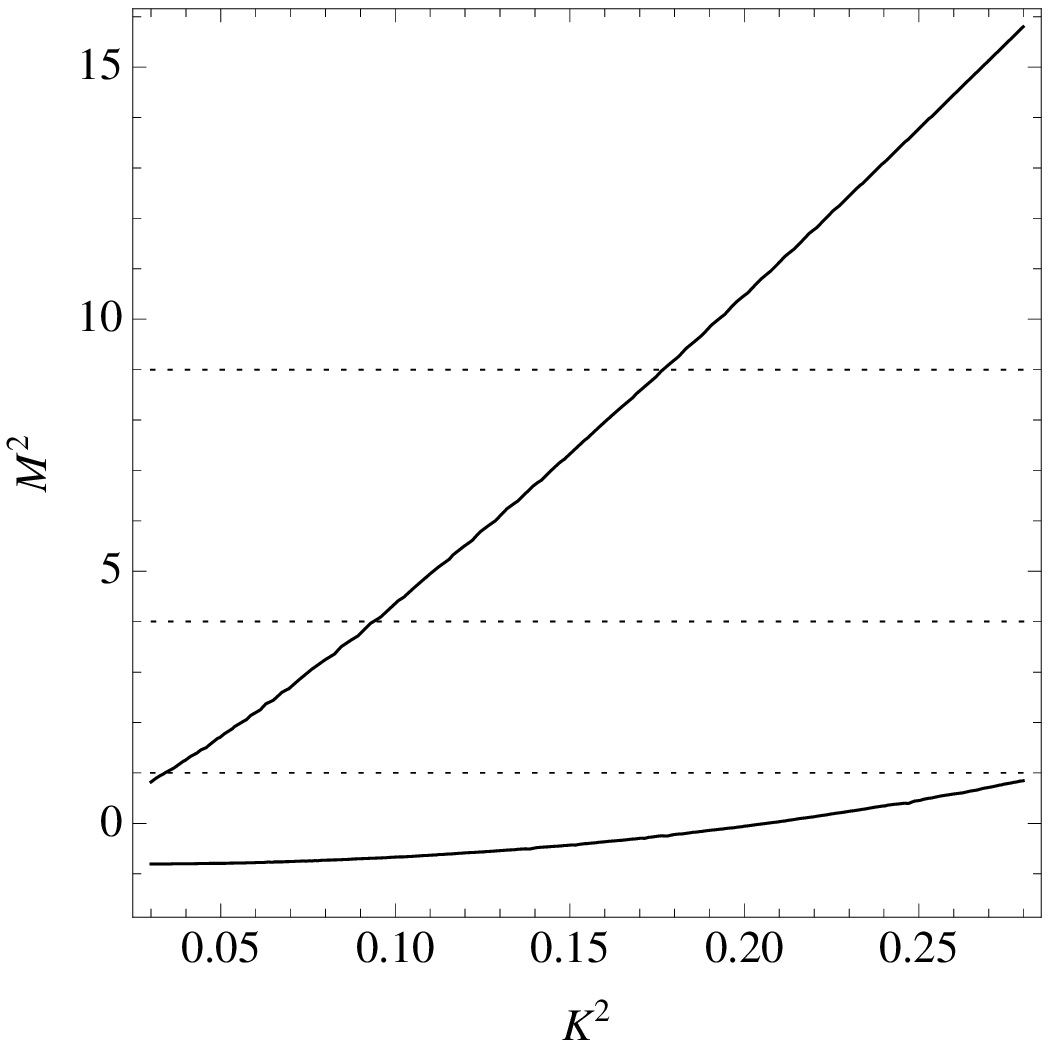}
\includegraphics[scale=0.55]{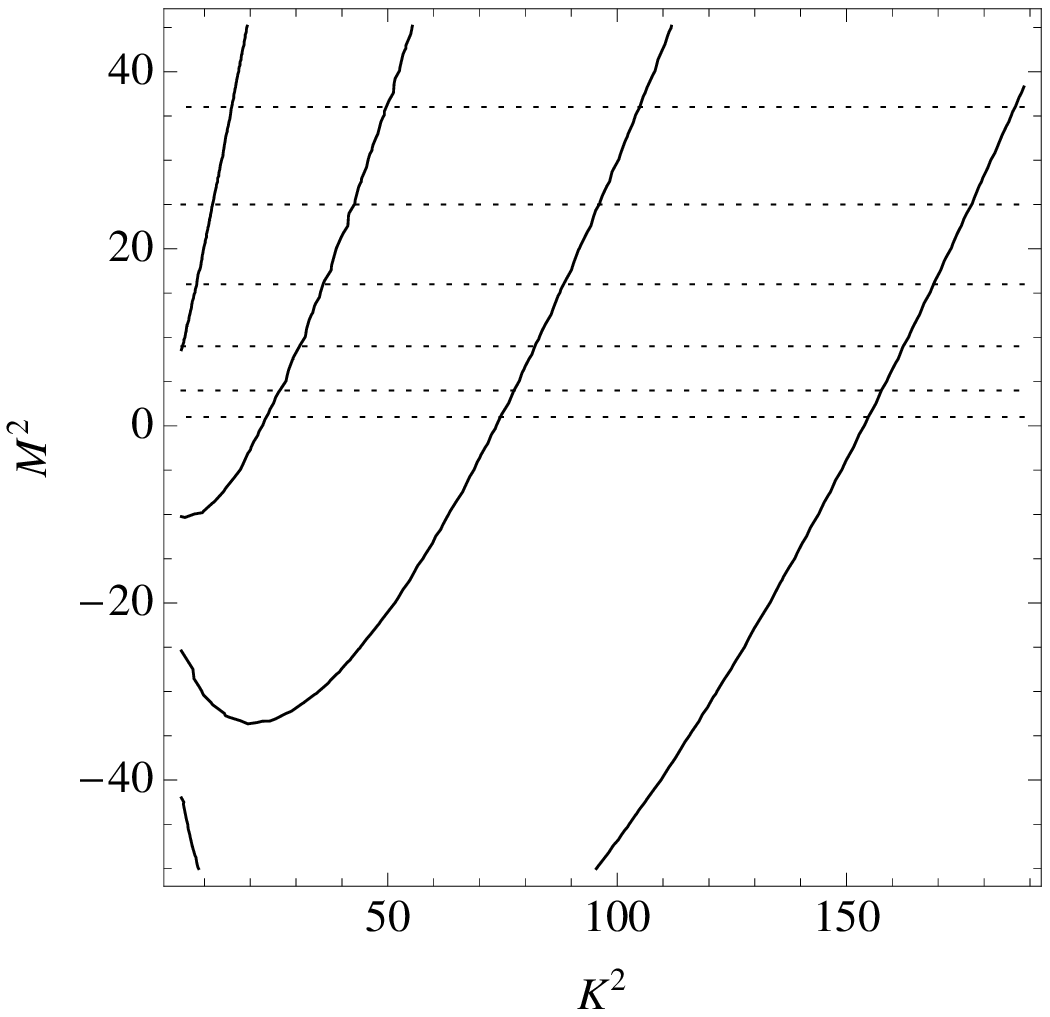}
\includegraphics[scale=0.55]{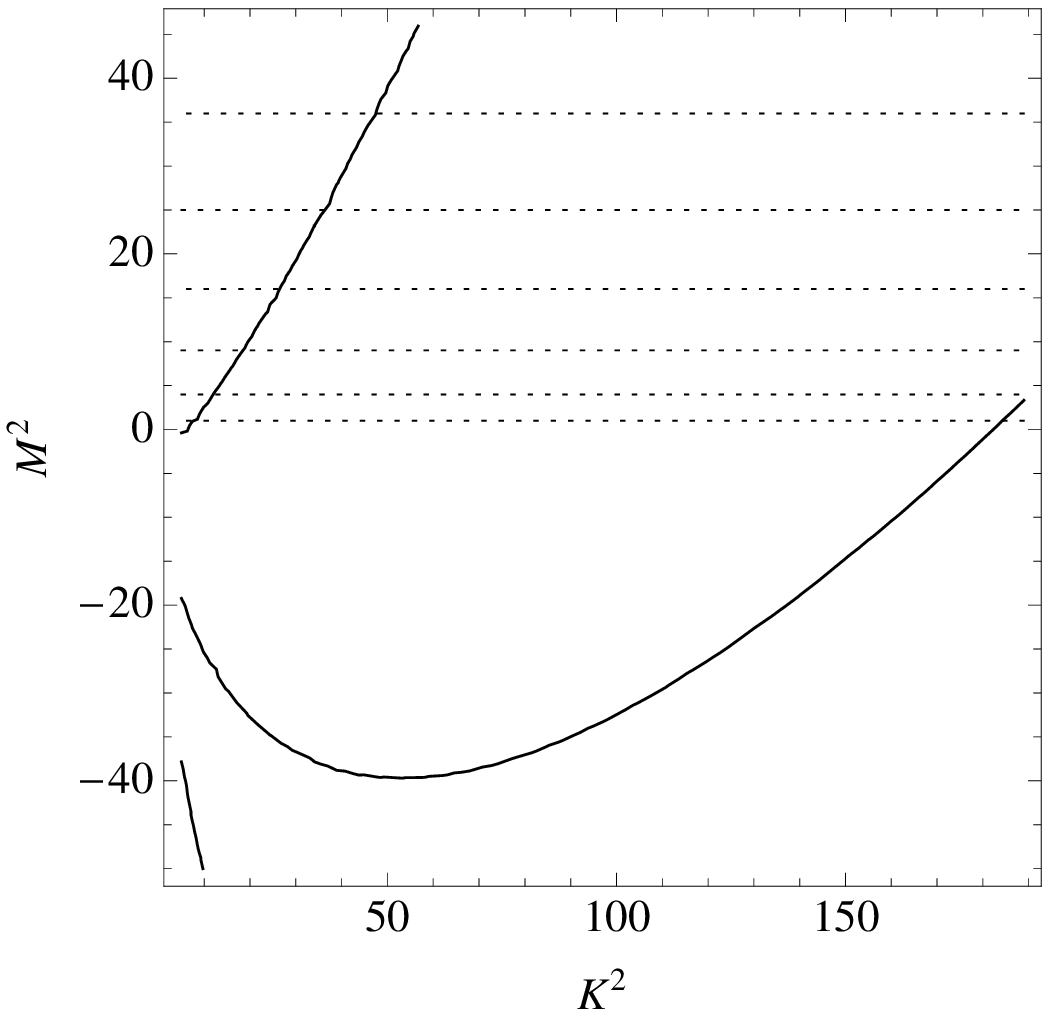}
\caption{Positions of the bound states together with distinguished lines $M\in\mathbb N$ for the Neumann boundary condition at $x_0=\{0.1,0.5,0.7\}$. All these plots are inequivalent in the sense, that they cannot be obtained from each other by a $\la-$transformation. }
\label{states}
\end{figure}

Let us remark on how figures such as Fig.\ref{states} are to be interpreted. Firstly, the value of $x_0$ is the specific characteristic distinguishing between the spectra (lines in the $(K^2,M^2)$ plane corresponding to bound states). Secondly, the $k=\sqrt{p^2-\w^2}$ will be known, once $x_0$, $r_0$ and $m$ are known, 
\begin{equation}
|k|=|m|  (\tfrac{x_0}{r_0})^2,
\end{equation}
which allows us to write 
\begin{equation}
K^2=m^2 (\tfrac{x_0}{r_0})^2. 
\end{equation}
This equation allows for multiple interpretations of a given point $(K^2,M^2)$ on the ``bound-state-line'', depending on the chosen value of $m\in\mathbb Z$. Thus, from given $(K^2,M^2,m)$ we compute the radius $r_0$, and the frequency $\w_m(M^2)$:
\begin{align*}
r_0 &=\frac{|m|x_0}{\sqrt{K^2}}, \\
\w_m(M^2)&=\jd(\tfrac{M^2}{m}-m).
\end{align*}
Furthermore, once $r_0$ is known, $|k|$, and (together with the known $\w_m$) $|p|=\sqrt{k^2+\w_m^2}$ can be computed.

 Points lying near $M^2=n^2$ ($n\in\mathbb N$) are distinguished, because they can correspond to zero-frequency modes. For instance, from the Fig. \ref{states}, with $x_0=0.1$ we find a bound state at \mbox{$M^2=1$}, $K^2\approx 0.034$, corresponding the zero-frequency, $\w_{1}(1)=0$, solution with $|m|=1$ and
\begin{equation}
r_0\approx 0.54, \qquad p=\pm (\tfrac{x_0}{r_0})^2\approx \pm 0.034.  
\end{equation}
Thus, in this case, it is a problem with an ergoregion, and the (essentially static) perturbation has a form of a helical structure, with weak $z-$dependence, localized close to the vortex ($\psi\sim K_{1}(p r)$ for large $r$). By considering states close to $M^2$, with the same $|m|$, we obtain problems (characterized by $r_0$'s) with similar solutions, but either co- (for $M^2>1$) or counter-rotating (for $M^2<1$) with the vortex.

 On the other hand we could have interpreted the above values of $(K^2,M^2)$ as coming from the second branch, $|m|=2$, for which we would obtain
\begin{equation}
r_0\approx 1.09, \qquad \w_{-2}(1)=0.75, \qquad    p=\pm \sqrt{\left[2(\tfrac{x_0}{r_0})^2\right]^2+\w^2} \approx \pm 0.75.
\end{equation}
This solution, therefore, corresponds to a problem without an ergoregion, and the perturbation (localized very close to the vortex) is counter-rotating (sgn$(\w m)<0$) and moving along the vortex. \\[2pt]

\subsection{ Location of bound states in the space of physical parameters; dispersion relations}\label{bound_dispersion}

The scaling transformations in the previous section allowed us to formulate ``existential'' statements to the effect that the existence of one bound state for certain values of physical parameters $r_0,\w,m,p$ was equivalent to the existence of bound states for other values of these parameters. From the physical point of view it is, however, of interest to determine also the full ``spectrum'' of solutions for a given boundary problem. In other words: for a given type of boundary condition at $r=0$, or at a finite $r=r_0$, we shall compute the lines corresponding to bound states in cuts of the the $(p,\w,m)$ parameter space. This is a rather large space of parameters, and below we will attempt to extract the essential characteristics of the solution of the problem, that is essential features of the three-dimensional surfaces embedded in the four-dimensional space $(r_0,p,\w,m)$.  We restrict this numerical consideration to the boundary conditions of Neumann type posed at a finite $r=r_0$. Only the case $\w>0$ will be presented, as the case $\w<0$ follows from  the symmetry $(\w,m)\leftrightarrow (-\w,-m)$ of the problem. The form of the dispersion relations is qualitatively different for co- and counter-rotating solutions, as defined by the sign of $m$.  (There are no bound states with $m=0$,  as the radial functions would need to be $K_0(kr)$, $k=\sqrt{p^2-\w^2}$, which however is a monotonic function and cannot fulfill the Neumann boundary condition at any $r_0$.)  

For the {\bf counter-rotating solutions ($m<0$)} we obtain  a number of branches concentrated just  below the diagonal $\w=p$ (see Fig. \ref{disp_counter}). Solutions generally exist for all values of $r_0$, but small frequencies are only possible in the presence of an ergoregion, i.e. if $r_0<1$. The group velocities approach (but are always less than) $1$, meaning that packets constructed out of the wavefunctions corresponding to bound states would move along the vortex with almost the speed of sound.  
\begin{figure}[h!]
\centering 
\includegraphics[scale=0.55]{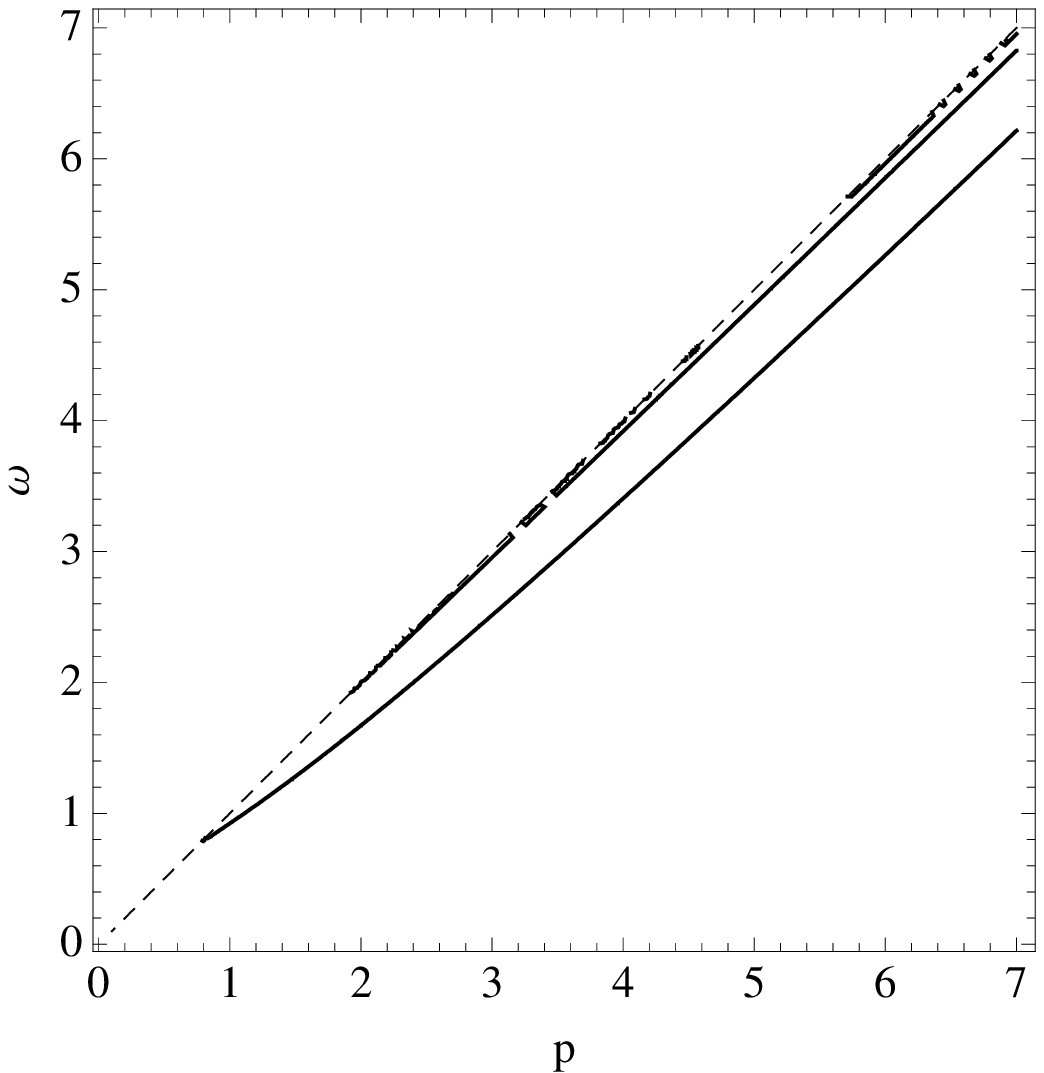}\qquad
\includegraphics[scale=0.59]{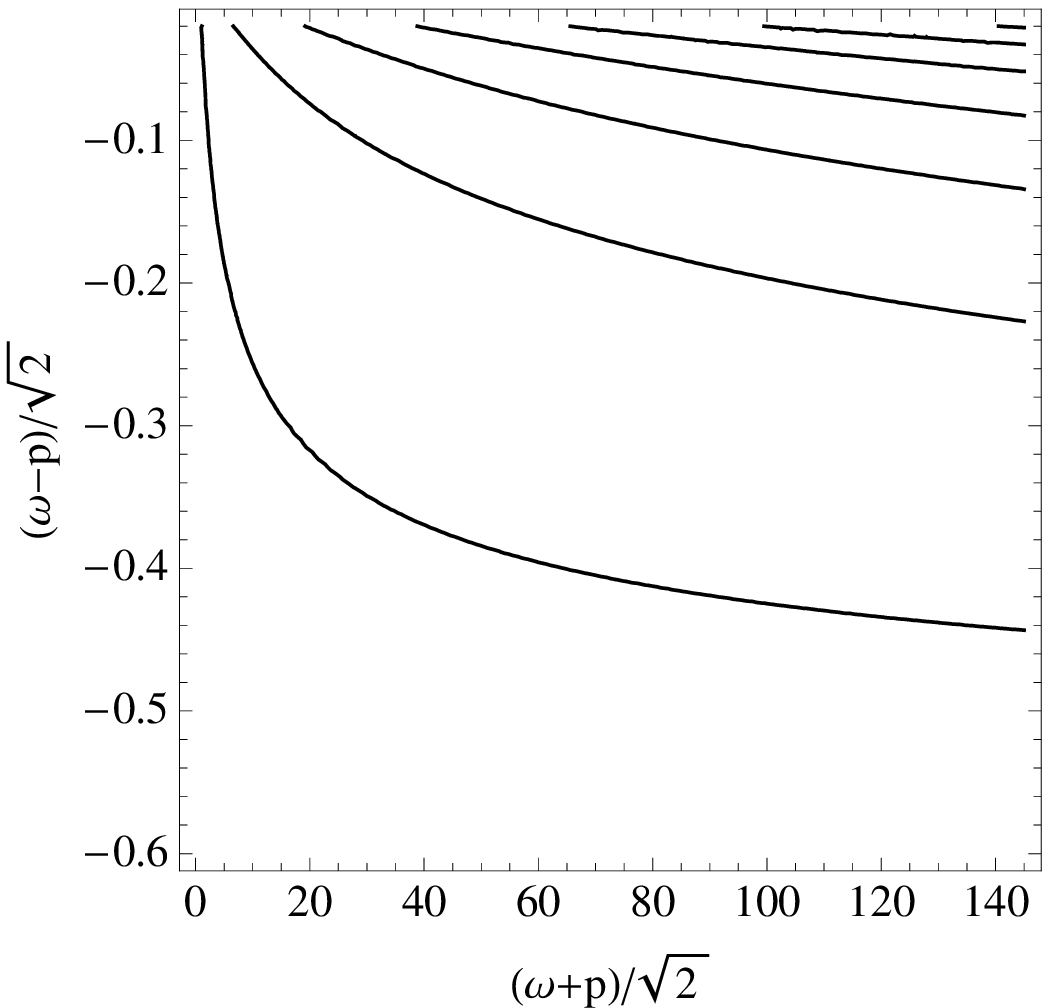}
\caption{[Left:] Typical ``band-structure'' for bound states localized close to the core, propagating along the vortex with a momentum $p$, and \u{counter-rotating} with it for $(r_0,m)=(1.09,-2)$. [Right:] Rotated by $\pi/2$ and extended view for the same values of $(r_0,m)$.}
\label{disp_counter}
\end{figure}
Given the observation, that each branch of solutions appears to begin at $\w=p$ we may ask: what is the lowest frequency $\w$ of a solution(s) corresponding to various $m$ for a given $r_0$. In Fig. \ref{lowest_omega} we have plotted the positions of the branches of bound states in the $(\w,m)$ plane (with $p=\w+0.002$). There is an apparent qualitative change between $r_0<1$ (ergoregion) and $r_0>1$ (no ergoregion): in the latter case no solutions with $\w\approx 0$ are possible. In the cases barely admitting ergoregions ($r_0\lessapprox 1$) the only branch approaching $\w=0$ is the leftmost one, and it does so for rather large $|m|$ (see the central plot of Fig. \ref{lowest_omega}). For $r_0>1$ the solutions with small $|m|$ are the ones leading to lowest $\w$'s, although not necessarily the $m=-1$ mode. These modes are all gapped, in contrast to the gapless Kelvin mode in the unpinned situation, \cite{Donnelly,Sipp2003,Isoshima}.

\begin{figure}[h!]
\centering 
\includegraphics[scale=0.51]{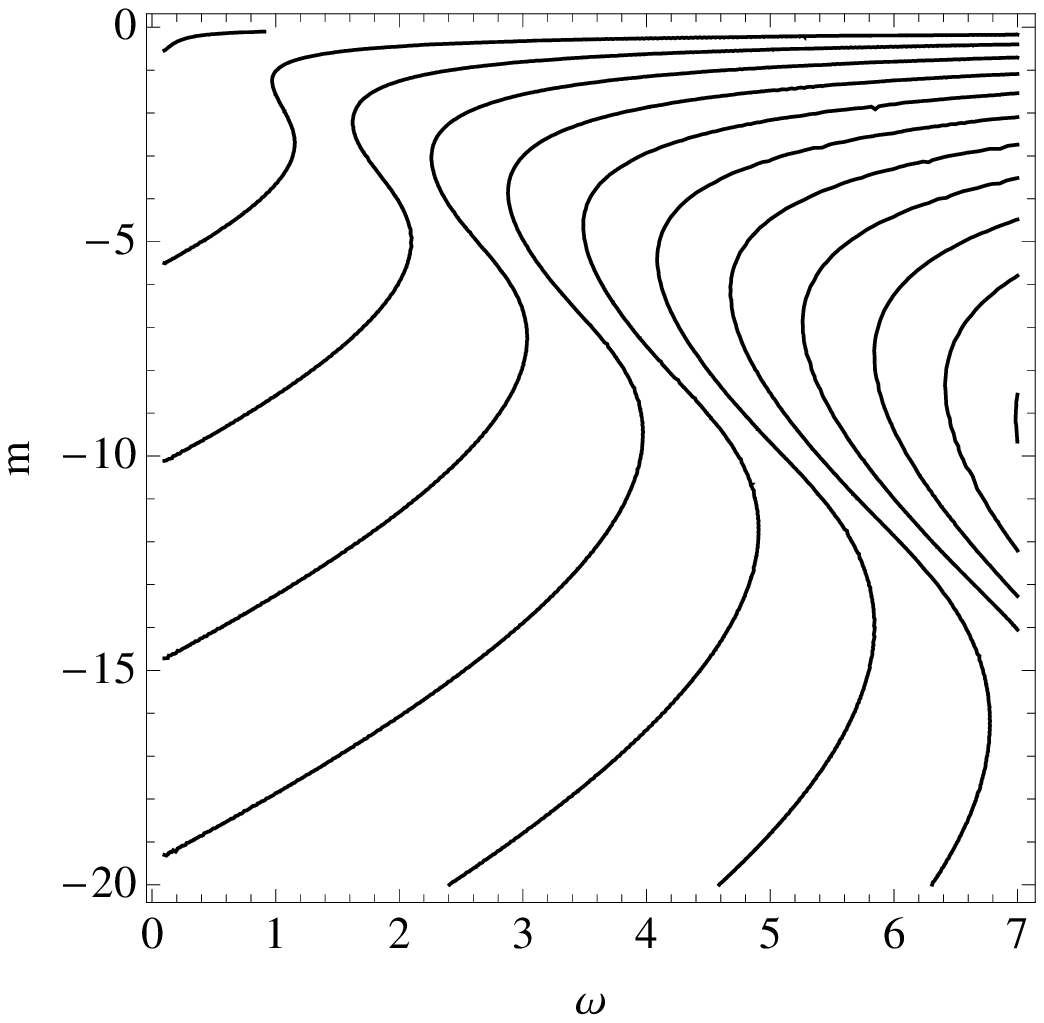}\qquad
\includegraphics[scale=0.51]{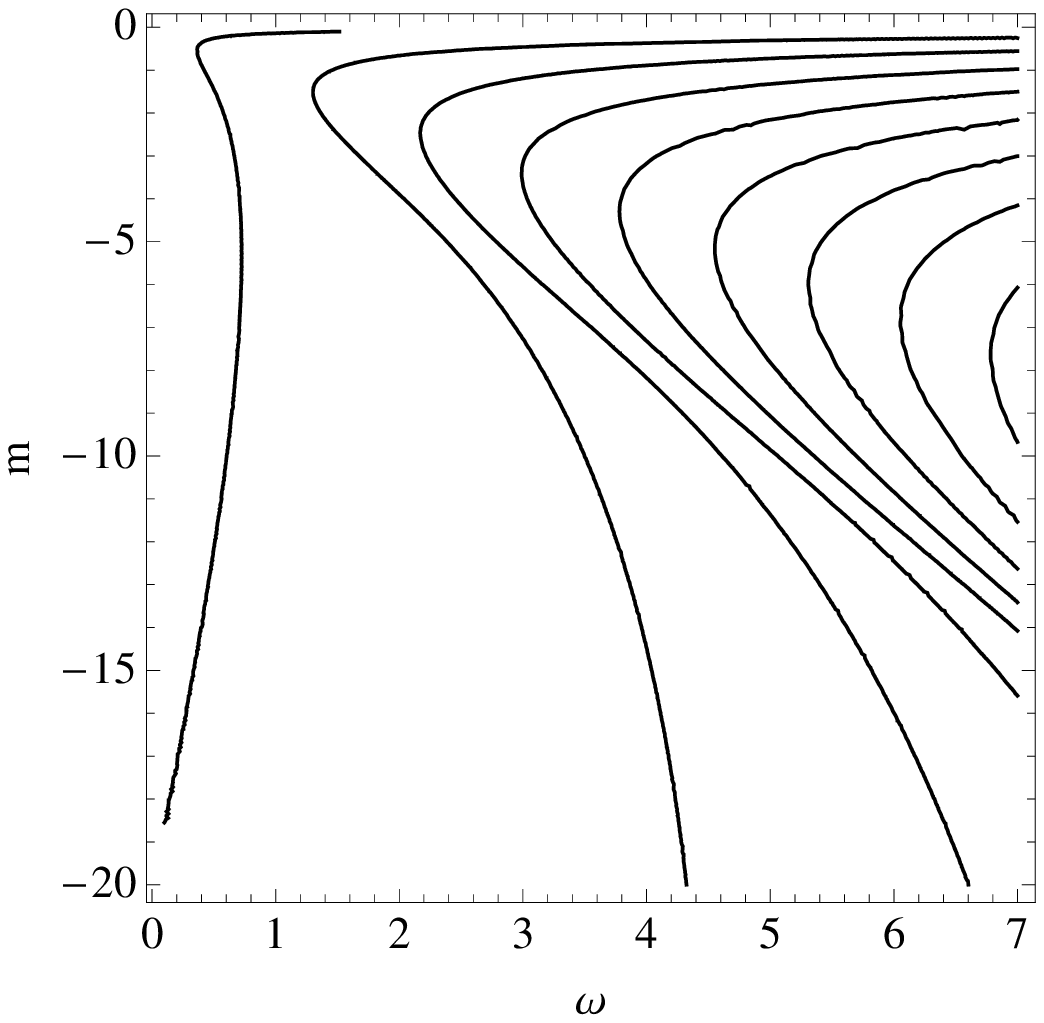}\qquad
\includegraphics[scale=0.51]{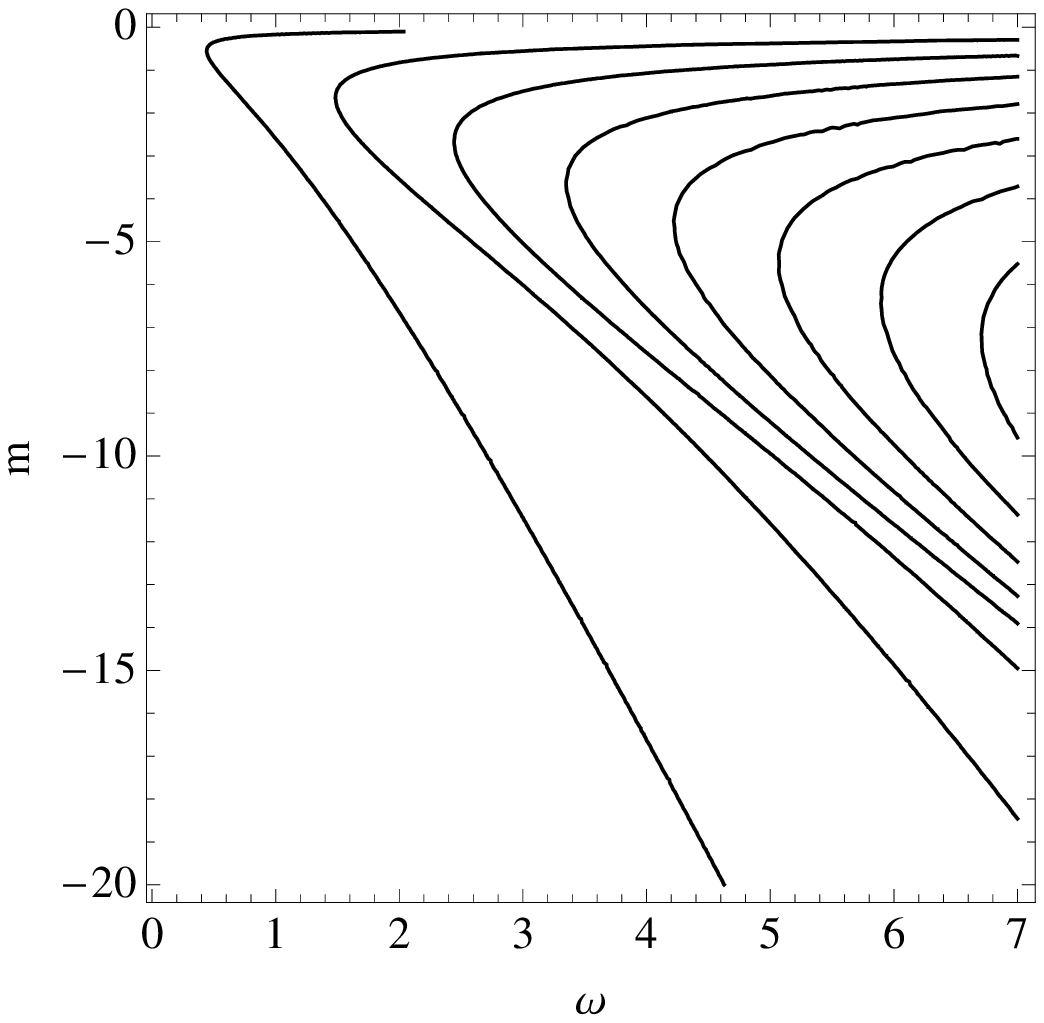}
\caption{Positions of counter-rotating bound states with $p=\w+0.002$ for $r_0=\{0.5,0.9,1.2\}$. Zero-frequency states exist only for $r_0<1$ and are related to leftmost branches in the plots. From $r_0>1$ up to $r_0\approx 8$ the state with $m=-1$ is the lowest-$\w$ state, while for larger $r_0$ this role is played by states with ever lower $m$. }
\label{lowest_omega}
\end{figure}

The {\bf co-rotating solutions ($m>0$)} exist only in the presence of an ergoregion. The branches are characterized by negative group velocities, all of which start at the diagonal, $\w=p$, and end on the line $\w=0$  (see  Fig. \ref{disp_co}). As the Neumann radius $r_0$ is increased (for a given $m$) the branches move towards small $p$, and some of them disappear. For each $m$ there exist a critical radius $r_0$, above which no co-rotating solutions exist. 
\begin{figure}[h!]
\centering 
\includegraphics[scale=0.50]{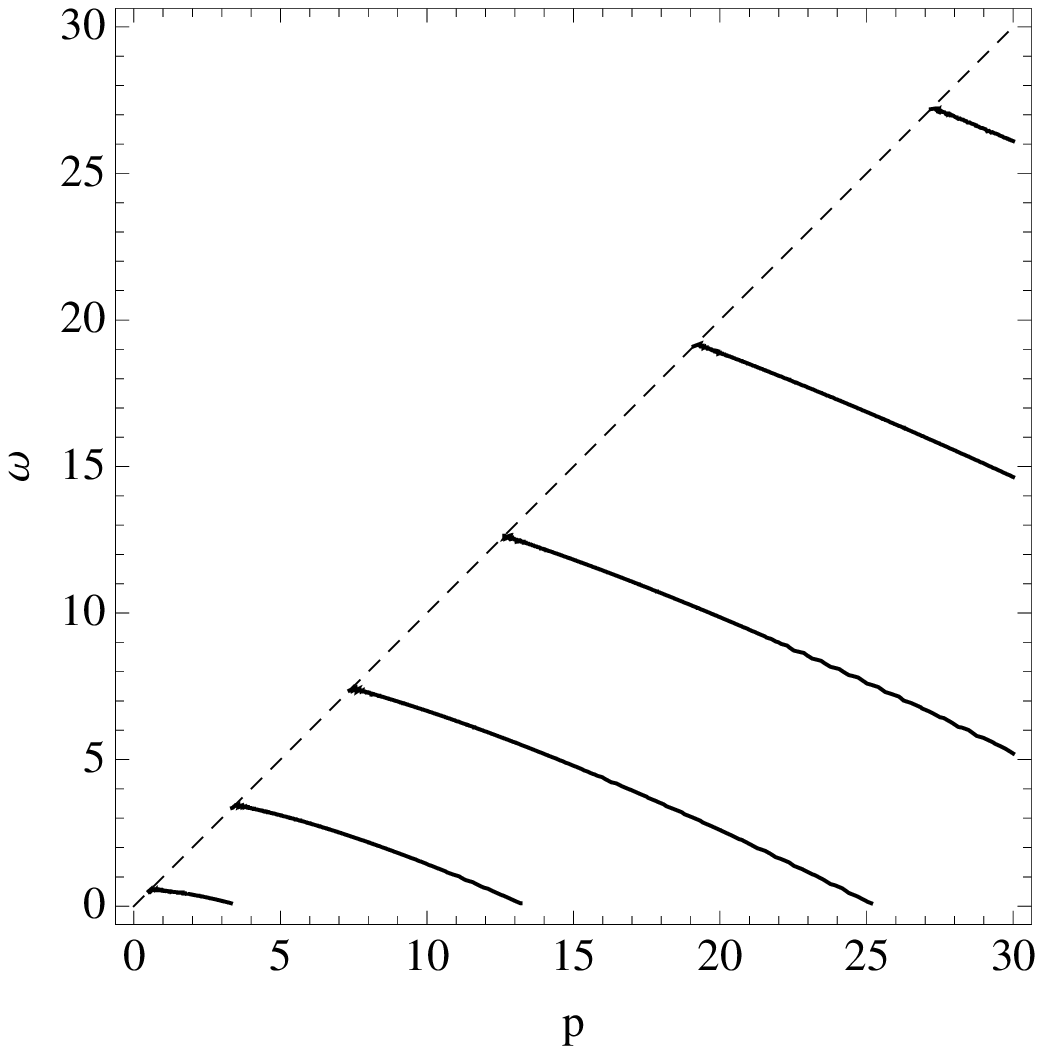}\qquad
\includegraphics[scale=0.50]{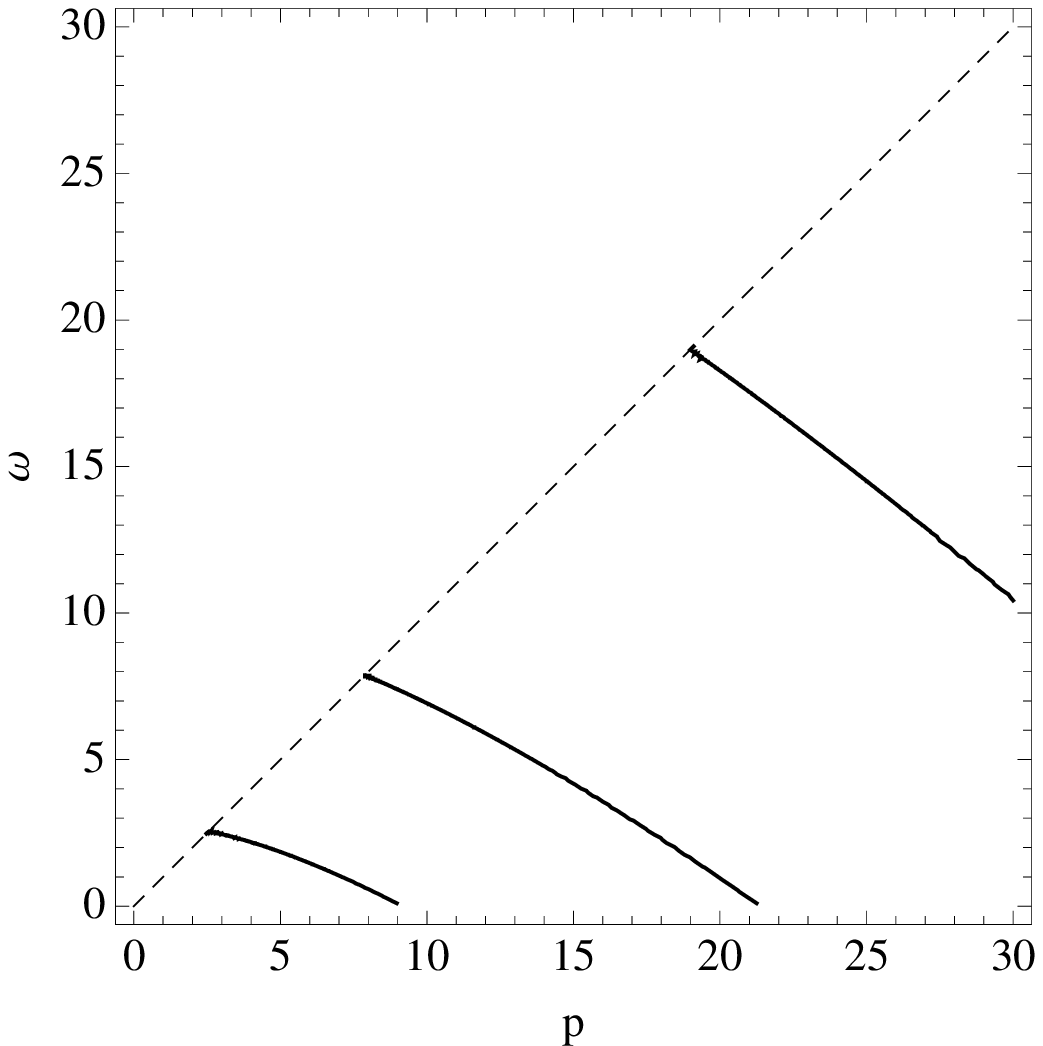}\qquad
\includegraphics[scale=0.50]{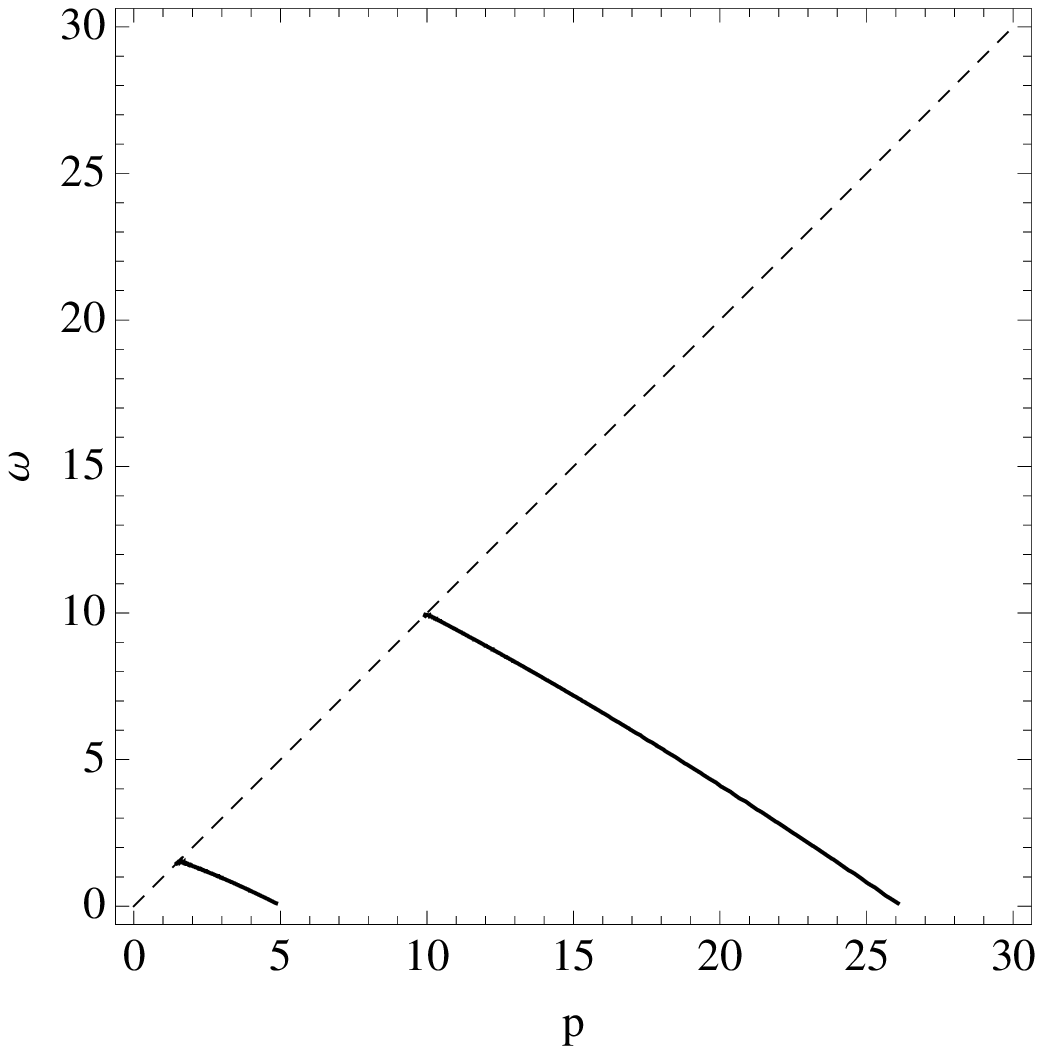}
\caption{Typical ``band-structure'' for bound states localized close to the core, propagating along the vortex with a momentum $p$, and \u{co-rotating} with it for [Left:] $(r_0,m)=(0.1,5)$, [Central:] $(r_0,m)=(0.3,5)$, [Right:] $(r_0,m)=(0.1,1)$. The plots correspond to a spacetime with an ergoregion; otherwise co-rotating bound states do not seem to exist.}
\label{disp_co}
\end{figure}
In the Fig. \ref{m_r_plot} we have depicted the positions of bound states in the $(r_0,m)$ plane, assuming fixed, small, values of frequencies and momenta, with $\w\approx p$. Given the way position of bound states moves as $r_0$ is increased (Fig. \ref{disp_co}) an appearance of state at $p\approx \w\approx 0$ signifies the entrance of a new branch into the $(p,\w)$ plot (dispersion relation). The rightmost line in Fig. \ref{mmm_r_plot} is the line of occurrence of the first co-rotating bound state, i.e. at a fixed $m$ no bound states exist if $r_0$ is greater than the position of this line. For $r_0$ just smaller than $1$, the co-rotating states exist only with very large angular numbers $m$.

\begin{figure}[h!]
\centering 
\includegraphics[scale=0.55]{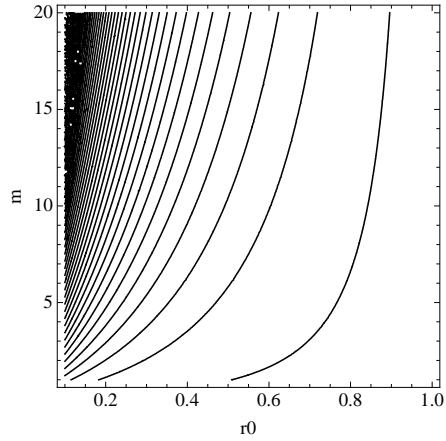}
\caption{Position of co-rotating bound states with small $\w$ and $p$ ($p=0.11$, $\w=0.1$) in the presence of an ergoregion as a function of the boundary radius $r_0$ and $m$. No bound states exist for $r_0>1$ (without ergoregion), and these that exist for large $r_0<1$ are necessarily associated with large $m$.}
\label{mmm_r_plot}
\end{figure}

\subsection{Stability of bound-state modes (complex frequencies), relation to Friedman (ergoregion-) instability}\label{vortex_stability}
In the fluid-dynamical context, by considering small (test) perturbations of a given flow one can distinguish two types of instabilities. The first type develops, when the global energy of the fluid is lowered upon introduction of some perturbation. The  second type comes about when modes of perturbation with complex eigenvalues are allowed, that is fulfill the appropriate wave equation and the boundary conditions. An example of the the first, milder, type of instability is provided by the creation of excitations in superfluid flows exceeding the Landau velocity \cite{LL3}. It is natural to expect this type of instability in the presence of ergoregions because of the non-positivity of the energy density of perturbations there. However, it is difficult to estimate how rapid the dynamics associated with the instability will be, as the rate of creation of (negative-energy) excitations depends on external factors such as the surface roughness for capillary superfluid flows. In the vortex case, one can easily think of situations, where the fluid will not encounter any factors efficiently generating negative-energy excitations. Invoking the spacetime analogy: it is difficult to estimate a priori the probabilities of Penrose-type processes, \cite{Poisson}.

Due to the relation \eqref{EnergyNormRelation} negative energies are associated with positive-frequency, negative-norm states\footnote{And, of course, by $(\w,m)\ra(-\w,-m)$ symmetry, with negative-frequency, positive-norm solutions.}. Naturally, this cannot happen when  $m<0$ (for counter-rotating modes), because the norm is explicitly positive definite in this case. For co-rotating ($m>0$) modes we can generally assess that at least the energies of the modes close to $\w=0$ will be negative. This is always possible (see Fig. \ref{disp_co}) whenever such modes exist, but as we have argued it requires the presence of an ergoregion (see e.g. Fig. \ref{mmm_r_plot}).

With regard to this first-type (negative-energy) of instability let us note the connection to the work of \pers{Friedman}, where it is argued, that astrophysical spacetimes with ergoregions but without horizons are unstable \cite{Friedman}. The theorem proved there states, that unless a test, massless field can settle down to a non-radiative time-dependent state, the energy of the system in the bounded vicinity of the object will tend to $-\infty$, radiating an infinite amount of energy to null infinity, provided the system starts from a negative energy state. The vortex spacetime provides a model of a spacetime where an analogous statement should also hold\footnote{\pers{Friedman}'s work is related to the astrophysical context, and therefore assumes asymptotic flatness. Although the vortex (acoustic) spacetime due to its cylindrical symmetry does not conform to this assumption, we just use it to make the point, that bound states would provide a way out of the conclusion of the theorem of \pers{Friedman}.}. Note however, that our co-rotating bound states provide exactly the type of solutions the the ``unless'' clause would exclude: they are time-dependent, possibly negative-energy, non-radiative solutions of the d'Alembert equation in a stationary axisymmetric spacetime with an ergoregion. We do not see how, in the astrophysical context, the existence of the bound-state-like 
states can be excluded, but such states, decaying exponentially at large $r$ (as in our vortex case) would violate the assumed asymptotic form of the solutions of the wave equation (Eq. (23)) in \cite{Friedman}.
\begin{figure}[h]
\centering 
\includegraphics[scale=0.50]{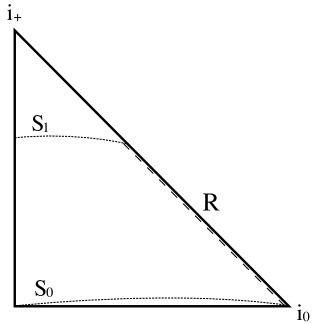}
\caption{Penrose diagram relevant to the theorem of \pers{Friedman}, \cite{Friedman}. Part of the energy of a field configuration set up on the Cauchy surface $S_0$ is radiated away through $R$, which belongs to the null infinity. Because the radiated energy is necessarily positive, the energy of the same field configuration on the (non-Cauchy) surface $S_1$ is lower.}
\label{m_r_plot}
\end{figure}
 Moreover, as the existence of bound states with negative energy in our work is seen to be related to the presence of the ergoregion, we doubt, that one can argue against the existence of such states (in any context) by examining fall-off conditions of the field at null infinity alone (as is done in \cite{Friedman}). If the completeness property for eigenmodes of the Hamiltonian turns out to hold in the ergoregion case, where it is a symmetric operator on a Krein space only, one will argue, that every wavepacket can be expanded into the eigenmodes, and the part associated with scattering states will be radiated away, carrying positive energy, while the part associated with bound-states will remain close to the innermost part of the system forever.\\[3pt]

 Let us finally comment on the second type of instability, associated with complex frequencies, which is typical of unstable  configurations of fluids. By the known properties of symmetric operators in Krein spaces (see appendix \ref{J-adjoint}) complex isolated eigenvalues of the Hamiltonian, \eqref{EOM}, always appear in pairs of mutually conjugate numbers. The radial part of the Hamiltonian and the boundary condition associated with it depend, however, on three free, real parameters:  $p$, $r_0$ or $m$. Let us call such an operator $H_\xi$, where for the purpose of the argument we investigate the dependence of the spectrum on one of the parameters, say $\xi$, with the other fixed\footnote{We also extend the domain of $m$ to real numbers.}. Relying on the analogy with similar (Hamiltonian, not necessarily self-adjoint-) problems \cite{SS,Fulling_book,KleinRafelski}, and on the experience with finite-dimensional operators, \cite{Baumgartel}, we expect that one signature of an appearance of complex frequencies would be if a pair of real frequencies (two lines in the $(\xi,\w)$ plane, say for $\xi>\xi^*$) were to ``merge'' at a point $(\xi^*,\w^*)$, where the tangent line would be vertical, $\left.\tfrac{d\w}{d \xi}\right|_{\xi^*}=\infty$. Beyond this point (i.e. for $\xi<\xi^*$) only two mutually conjugate complex frequencies would remain.  In the Fig. \ref{instab} we present the dependence of $\w$ on various choices of $\xi$ in what we observe to be the general behavior: no points of vertical tangent are to be found in the spectrum. We therefore conjecture that no complex frequency modes ever develop from a merger of real frequencies (in the vortex problem). However, the other way for complex frequencies to enter the spectrum is for them to enter from the continuous  spectrum, i.e. from the region $|\w|>|p|$. Such an entry would leave no trace on plots of the positions of bound states in $(\xi,\w)$ planes (because on these plots $\w\in \mathbb R$ per assumption), and requires further investigation.

\begin{figure}[h!]
\centering
\includegraphics[scale=0.50]{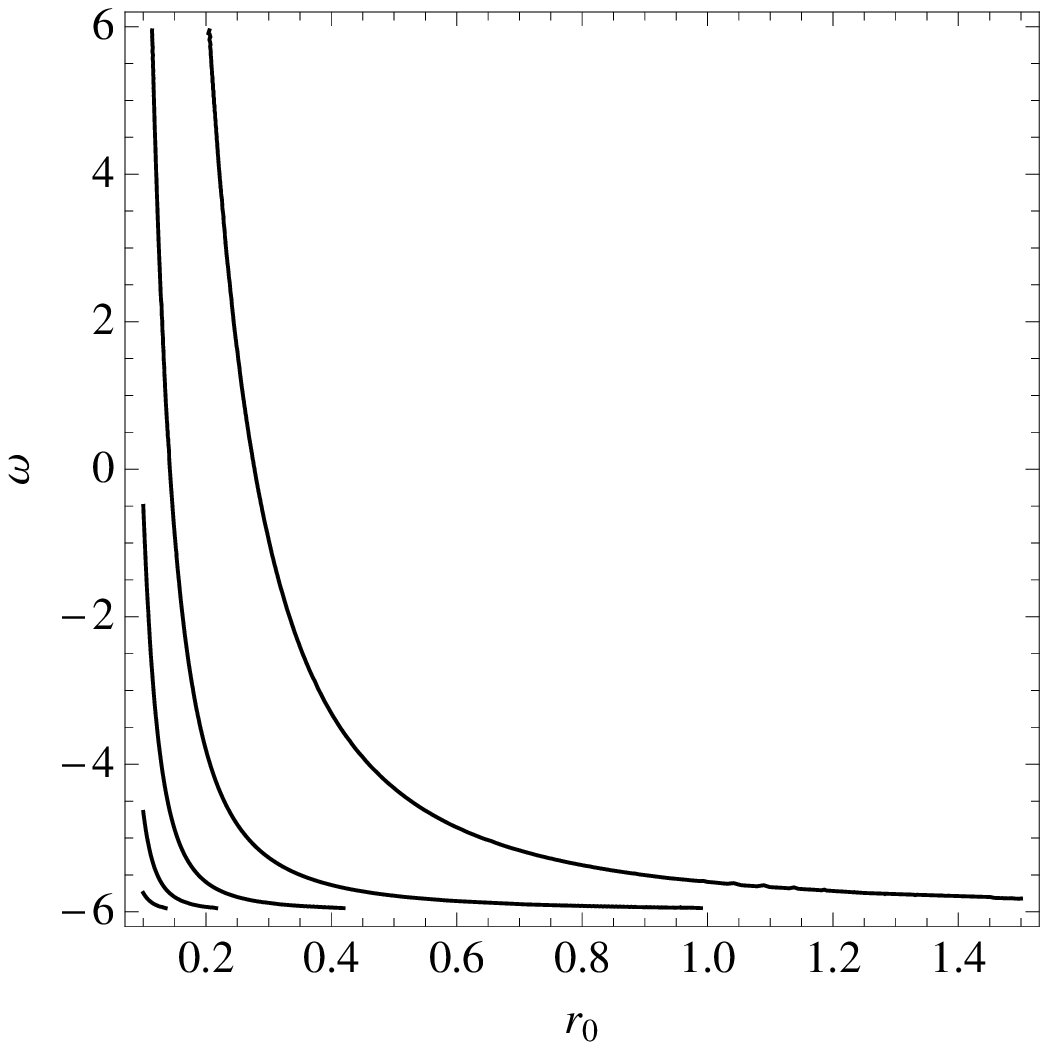} \qquad
\includegraphics[scale=0.50]{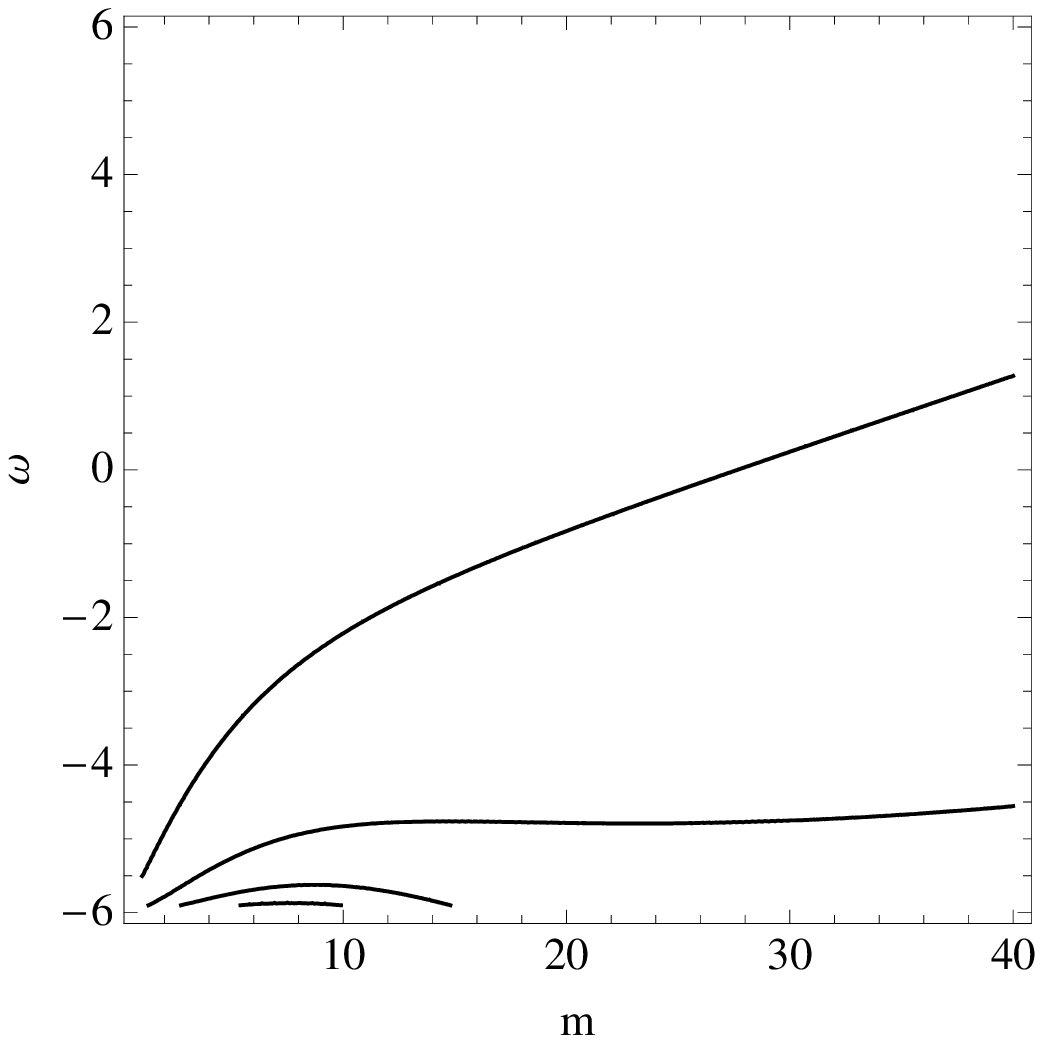} \qquad
\includegraphics[scale=0.50]{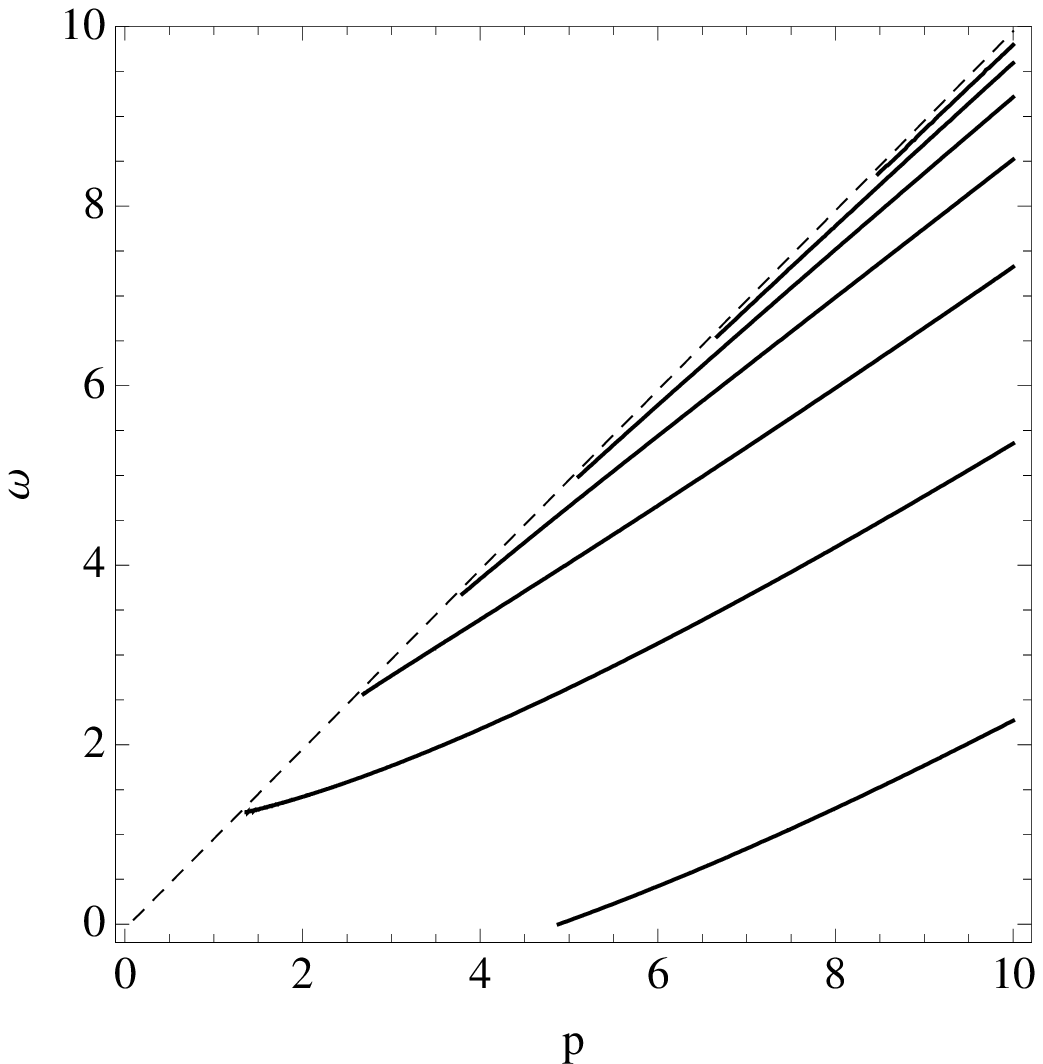} \qquad
\caption{Typical positions of bound states: [Left:] in the $(r_0,\w)$ plane (Neumann problem) for fixed $(m,p)=(1,6)$ [Center:] in the $(m,\w)$ plane for fixed $(r_0,p)=(0.9,6)$ [Right:] in the $(p,\w)$ plane for fixed $(r_0,m)=(0.2,-5)$.}\label{instab}
\end{figure}

In order to give an example of how complex frequencies can manifest themselves in problems involving sound propagating on rotating background flows we recall the result of Sipp and Jacquin \cite{Sipp2003}. In this work an inviscid fluid forming a Lamb-Oseen vortex, $\Gamma(r)= \frac{const}{r}\left(1-e^{-r^2/r_0^2}\right)$, is investigated and it is found that many of the Kelvin waves are \emph{damped}, i.e. possess frequencies with negative imaginary parts. Moreover, as one moves along a \emph{single branch} (mode) in the $(p,\w)$ plane with an (initially) real frequency $\w$, a non-zero damping rate suddenly emerges at some critical value of $p$. The relation of this result to our above argument is however not straightforward, because Lamb-Oseen vortex involves vorticity of the background flow and cannot be handled by the simple acoustic-spacetime description\footnote{An investigation, comparable to the acoustic-spacetime description, would need to use the extended formalism of \cite{PerezBergliaffa2004121}. While this, to the best of our knowledge, has not been attempted yet, it would provide an interesting extension of the work of Flaig and Fischer, \cite{FischerFlaig}.}. This description leads to the structure with  $H$ being a symmetric operator on a Krein space (see chapter \ref{lin_dyn_sys}).


\section{Rotating superfluid flows accessible to experiments}\label{experiments}

Here we will recall the commonly available experimental setups, and judge their relation to the problem investigated in this paper.  Attention will be paid to supersaturated quantum vapors (BECs) and to the superfluid helium-4.

\subsection{Supersaturated quantum vapors}
\subsubsection*{Scattering length, mass, interaction strength, vortex circulation for \Rb}
Common experiments with rotating BECs (see e.g. \cite{Fetter_Review,BretinPhD}) use doubly spin-polarized $^{87}$Rb atoms. The relevant scattering length in this case is the the triplet one\footnote{$a_0=5\cdot 10^{-9}$cm is the Bohr radius.} $a_t=105\pm 4 a_0$, that is $a_t=5\cdot 10^{-7}$cm. Using the atomic mass of\footnote{We denote the mass of the boson constituent of the condensate by $m_B$.} \Rb, $m_B=87u=1.4\cdot 10^{-22}$g we estimate the interaction strength
\begin{equation}
g=\frac{4\pi\hbar^2a}{m_B}=4.5\cdot 10^{-38}\ \frac{\text{g cm$^5$}}{\text{s}^2}\,  
\end{equation}
The quantum of circulation is $\Gamma_0=\int v_idx^i=2\pi v_\vp=\frac{2\pi \hbar}{m_B}=4.5\cdot 10^{-5}\ \frac{\text{cm}^2}{\text{s}}$. In the acoustic metric there appears 
$$\kappa=v_\vp=7\cdot 10^{-6}\,\frac{\cm^2}{\sek}.$$

\subsubsection*{(Some) Experimental values of $n$, $c_s$, $\xi$ and $r_c$}
Typical contemporary experiments involve  condensate clouds with $N=3\cdot 10^5$ atoms in a cigar-shape of characteristic dimensions $10^2\times 100\, \mu$m$^3=10^{-8}$cm$^3$ corresponding to a typical number density of $n=3\cdot 10^{13}\cm^{-3}$. The typical speeds of sound are therefore of the order
\begin{equation}
c_s=\sqrt{\frac{\d p}{\d \rho}}=\sqrt{\frac{g n}{m_B}}=10^{-1}\frac{\cm}{\sek}. 
\end{equation}
For vortices we get the critical radius (radius of the ``ergoregion'')
\begin{equation}
r_c=\frac{\kappa}{c_s}=7\cdot 10^{-5}\cm=0.7\mu\meter, 
\end{equation}
which is exactly the coherence length (below which the density and $c_s$ begin to vary with $r$) multiplied with $\sqrt{2}$, 
\begin{equation}
\xi=\hbar/\sqrt{2m_Bgn}=r_c/\sqrt{2}=0.5\mu\meter.
\end{equation}
The hydrodynamic description of BECs becomes inadequate on distances comparable  with $\xi$, and therefore even if the the core of the vortex could be approximately described by boundary condition of one of the types considered in this work, it would still  correspond to an acoustic spacetimes without ergoregions. As a consequence of the relation $r_c=\sqrt{2}\xi$, conceivable modifications of setups with a BEC-vortices (involving changes to number density $n$, interaction strength $g$, mass/type of the boson $m_B$) would not change this state of affaris. On the other hand, the issue is open for pinned BEC-vortices or vortices in condensates trapped by anharmonic external fields due to the modification of vortex structure in such setups, \cite{Isoshima,Baym}. Namely, e.g. for condensates trapped by potential growing quicker than $r^2$ (or for condensates trapped by  cylinders with solid walls) a new picture arises \cite{Baym}. The regular Tkachenko lattice of vortices begins to get replaced in the center by an empty core (giant vortex), where condensate density is zero. The velocity circulation around this core is finite, but the velocity of sound becomes negligible due to the vanishing of condensate density. Even though there are (usually) still vortices in the bulk, and the density is far from being constant everywhere\footnote{Density varies according to $n(r)\sim r^2-R_h^2$, speed of sound as $c_s\sim \sqrt{n(r)}$. For highest angular momenta, as the giant vortices begins to emerge, there remains a fraction of the Tkachenko lattice for $r>R_h$ with a constant lattice spacing $\ell$ (see \cite{Baym}); the velocity field corresponding to the missing lattice (for $r<R_h$) still circulates around the core of the giant vortex.}, the system exhibits characteristics of a spacetime with an ergoregion.

\subsection{Superfluid helium-4}
\subsubsection*{Single vortices}
The typical speed of (first) sound in the superfluid helium-4 is $c_s=2.4\times 10^4\frac{\cm}{\sek}$. Together with the mass of $^4$He atom, $m_{\He}\approx 4u=6.64\times 10^{-24}$g, this corresponds to $\ka=\frac{\Gamma_0}{2\pi}=1.588\times 10^{-4}\tfrac{\cm^2}{\sek}$, where $\Gamma_0=\tfrac{h}{m_\He}$ is the quantum of circulation; with these values we obtain
\begin{equation}
r_c=0.66\times 10^{-8}\cm
\end{equation}
which is \emph{less than the van der Waals radius} of the Helium atom, $r_{\rm vdW}\approx 1.4\cdot 10^{-8}\cm$. Consequently, single vortices in superfluid helium-4, when pinned on a wire to restrict their position, provide an arena relevant to this work, but with the Neumann (divided by $r_c$) greater than $1$ (no ergoregion).

\subsubsection*{Structure of a multiply charged vortex line; ratio $v/c$}
One would hope to create a model of ergoregion just by increasing the rotation of a vortex pinned on a wire. Let us contemplate this possibility for helium-4. General features of the structure of  vortex lines can be inferred from the empirical, $T=0$, equation of state (EOS) published  long time ago by \pers{Brooks} and \pers{Donnelly}, \cite{Brooks_Donnelly}:
\begin{equation}
p=\sum_{n=1}^3 A_n (\rho-\rho_0)^n,
\end{equation}
with $A_0=560$, $A_1=1.097\times 10^4$, $A_2=7.33\times 10^4$, (with units giving pressure in atm), where $\rho_0=0.145$ g/cm$^3$ is the density at the saturated vapor pressure (SVP). The equation gives approximately correct compressibilities as well as the Gr\"uneisen constant $U_G=\tfrac{\rho}{p}\tfrac{dp}{d\rho}|_T$, up to the pressure where helium solidifies $p_{\rm max}\approx 25$atm. Below we present the structure of the vortex based on this EOS, in order to give the reader a rough estimate of how realistic the assumption $\rho=$const, which we have used in this work, is.  Because the fluid at $T=0$ can only exist in the superfluid phase for pressures between SVP and $p_{\rm max}$ the density is only allowed to be in the range $[\rho_0,1.188\rho_0]$. From the EOS we compute the enthalpy density, $w(p)=\int^p\! \frac{d\tilde p}{\rho(\tilde p)}$,  which for $T=0$ can also be regarded as a function of $\rho$, and is a strictly increasing function. The vortex profile, i.e. $\rho=\rho(r)$, is obtained from the Bernoulli equation:
\begin{equation}
w(\rho(r))=w(\rho_\infty)-\frac{[v(r)]^2}{2}, \qquad \text{with $v(r)= \frac{\kappa}{r}$.}
\end{equation}
\begin{figure}[h!]
\centering 
\includegraphics[scale=0.7]{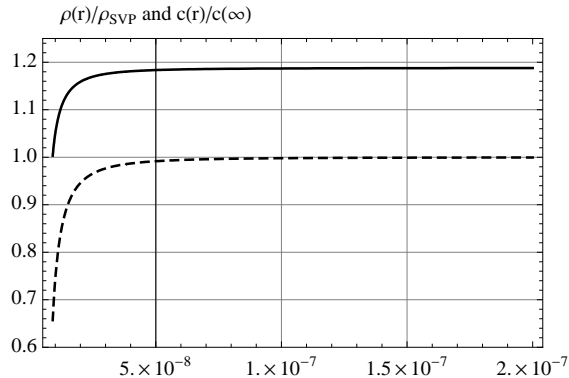}
\caption{The profile of density (solid) and speed of sound (dashed) for a vortex in superfluid helium-4 modeled with the equation of state of \cite{Brooks_Donnelly}. The chosen pressure at infinity is the maximal allowed pressure, and the vortex carries the single circulation quantum. Density is shown w.r.t. the saturated-vapor-pressure density, $\rho_{\rm SVP}=\rho_0$. Distances are given in [cm], and the innermost part should not be taken literally, because it is of the size of the van der Waals radius for $^4$He.}
\label{vortex_plot}
\end{figure}
The density (and thus enthalpy) decrease towards the center, and a boundary of the fluid develops at $r=r_*$, with  $\rho(r_*)=\rho_0$. In case we neglected the non-linear terms in the EOS (i.e. had we assumed a constant speed of sound), we would have obtained $\rho(r)=\rho_\infty \exp\left[-\tfrac{\kappa^2}{2c^2_\infty r^2}\right]$,
where we recognize the dimensionless combination $\tfrac{1}{2} (\tfrac{r_c}{r})^2$ in the exponent. Let us call such a case the simplified model. Generally, the speed of sound decreases towards the core of the vortex, and therefore the ratio $\tfrac{v(r)}{c(r)}$ grows monotonically. It is of interest to determine the highest possible value of this parameter; a value greater than $1$ would signify the presence of an ergoregion. Let $\rho_\infty=x \rho_0$; in the simplified model we find
\begin{equation}
r_*=\frac{r_c}{\sqrt{2\log x}}.
\end{equation}
While this result is valid for any fluid with the simplified equation of state $p(\rho)=c^2_\infty \rho$ ($c_\infty$ a constant), it is curious, that the maximal available $x$ for superfluid helium is $1.188$, making $r_*\approx 1.7 r_c$. Therefore, even though the size of the vortex, $r_*$, can be increased to avoid the difficulty with the van der Waals radius by increasing the circulation $\kappa$ (that is: by trapping more vortices on the central wire), the ratio $v(r)/c(r)$ will never exceed $1/1.7\approx 0.59$. No higher value of the ratio can be produced with superfluid helium not contained by external walls, within the simplified model (including the dependence of $c$ on $\rho$ does not change this conclusion significantly).

If, on the other hand, the fluid were contained by an external cylinder of radius $R$ (at pressure $\leqslant 25$atm), it would be possible to have almost arbitrarily large $v/c$. The values $r_*\leqslant r_c$ are achieved, however, only for relatively small diameters $R$; for instance in order to get $r_*=r_c$ one would need to take\footnote{Without the exterior container, this value of $x$ would require $\rho_\infty=212$atm (using unjustified extrapolation of the equation of state).} $x=1.65$ leading to $R=2.359\times 10^{-5}$cm and  $r_*=2.175\times 10^{-5}$cm. Even though both of these radii can be appropriately scaled up by choosing a larger circulation $\kappa$, $R$ will always remain only by $8\%$ larger than $r_*$. 

The  saturated-vapor-pressure density $\rho_0$, as well as the maximal density at which the superfluid remains liquid, $x\rho_0$, are clearly thermodynamic properties of the substance under consideration. It is difficult to give a principal reason why their ratio should not lead to $1/\sqrt{2\log x}<1$, although this is not the case in superfluid helium-4.

 \acknowledgements
 The author would like to thank N. Szpak, R. Sch\"utzhold and G. Jungman for simulating discussions. Financial support of the Deutsche Forschungsgemeinschaft (MA4851/1-1) is gratefully acknowledged. Initial part of this work was completed at the Leipzig University.

\appendix
 

\section{Spectral properties of selfadjoint operators in Krein spaces}\label{J-adjoint}
Let $H$ be a selfadjoint operator on a Krein space with the indefinite product $(.,.)$ (we drop the index $\KK$ here). The following properties hold (see \cite{Azizov,Langer,SS}):
\begin{enumerate}
\item[a)] if $\ba \w\neq \w'$ then the corresponding eigenvectors are orthogonal, $(\Psi_\w,\Psi_{\w'})=0$,
\item[b)] for each eigenvalue $\w$ there also exists an eigenvector to the eigenvalue $\ba \w$,
\item[c)] if an eigenvalue $\w$ is not real, then the corresponding eigenvector is of zero norm, $(\Psi_\w,\Psi_\w)=0$,
\item[d)] if to a certain eigenvalue $\w$ there also exists an associated eigenvector, $H\Psi^a_\w=\w \Psi^a_\w +\Psi_\w$, (where $\Psi_\w$ is the normal eigenvector to $\w$), then the normal eigenvector is of zero norm, $(\Psi_\w,\Psi_\w)=0$.
\end{enumerate}
According to \cite{SS}, in the case considered there (electrostatic potential $V(\vec x)$), the eigenfunctions together with the associated eigenfunctions form a complete set, so that any element of the phase space can be decomposed as (see eq. (3.33) of \cite{Fulling_book})
\begin{equation}
\Phi(t)=\sum_i\left\{ a_i\Psi_{\w_i}e^{-i\w_i t} +b_i\left[\Psi^a_{\w_i}-it\Psi_{\w_i}\right]e^{-i\w_it}\right\}, \quad \text{with }\Psi_\w=\Psi_\w(\vec x)
\end{equation}

\section{Positivity of the energy functional}
We introduce the decomposition $g_{ab}=-N_aN_b+e^\Phi_ae^\Phi_b+\g_{ab}$ and $T_a=\a N_a+ \be e^\Phi_a$, (for vortex spacetime $\be=-\tfrac{1}{r}$, and $\a=1$), and examine the integrand of the energy-functional $E[\psi]$. Using the notation: $\d_N\equiv N^a\d_a$, $\d_\Phi\equiv e_\Phi^a\d_a$, $(\d_\g\psi)^2\equiv \g^{ab}\d_a\psi\d_b\psi$, we find that
\begin{equation}
T_{ab}N^aT^b=\frac{\a}{2}\,\left[(\d_N\psi)^2+2\tfrac{\be}{\a}(\d_N\psi)(\d_\Phi\psi)+(\d_\Phi\psi)^2+(\d_\g\psi)^2 \right], 
\end{equation}
which is positive only as long as $|\be/\a|<1$. As this condition is not fulfilled in the ergoregion (in our case: for $r<1$), the density which is integrated over to obtain the energy $E[\psi]$ can be made negative for some classical field configurations.

\section{Relation to general theory of Bloch/Floquet/Mathieu} \label{mathieu_sec}
The equation 
\begin{equation}
\left[D^2_x-\tfrac{M^2}{x^2}+K^2(1+\tfrac{1}{x^4}) \right]\psi=0
\end{equation}
upon substitution $u=\ln x$, $u\in[-\infty,\infty]$ becomes the standard modified Mathieu equation
\begin{equation}
\psi_{uu}-[M^2-K^2\cosh(2u)]\psi=0,
\end{equation}
which in turn results from the Mathieu equation
\begin{equation}
\psi_{zz}+[M^2-K^2\cos(2z)]\psi=0
\end{equation}
taken along the pure imaginary axis $z=iu$. In the literature \cite{Abramowitz} the following definitions of constants in the Mathieu equations are employed:
\begin{equation}
M^2\equiv a, \qquad K^2=q.
\end{equation}
This is especially convenient for the reason, that the bound state problem (which appears in case there stands a $-1$ instead of $+1$ in the regular bracket of the original equation) is just the original problem taken along the line(s) $u=w\pm i\tfrac{\pi}{4}$, and with the parameter $q=\pm i K^2$ (the $x-$equation fulfilled by the bound state is recovered by substitution $x=e^w$). Our study  of solutions of the radial equation in the space of essential parameters is thus seen equivalent to the study of the solutions of the Mathieu equation:
\begin{itemize}
\item for real and positive $a$'s and $q$'s (corresponding to positive $M^2$ and $K^2$ -- scattering states for non-absorbed rays), for $z=iu$
\item for real $a$'s and $q$'s with a negative $a$ (corresponding to $M^2=-\mu^2<0$ -- scattering of ``absorbed rays'') $z=iu$
\item for real $a$ and purely imaginary $q$ (bound states) with $z=iu\pm i\tfrac{\pi}{4}$. The wavefunctions need to vanish at $x\ra+\infty$, which corresponds to Im$(z)\ra+\infty$.  
\end{itemize}

\begin{figure}[htb]
\centering
\includegraphics{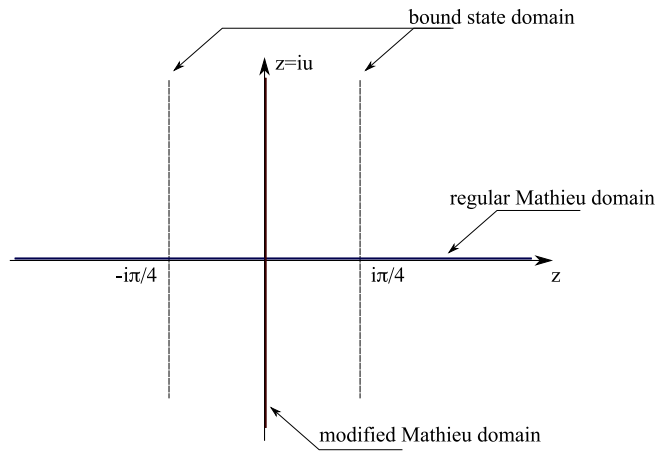}
\caption{Relation between Mathieu functions, and solutions of the radial equation for real frequencies $\w$. The region $z\ra i\infty$, corresponding to $x\ra \infty$, is where the bound state functions should exponentially vanish.}
\end{figure}

The double eigenvalue problems associated with the regular or modified Mathieu equation possess a number of special distinguished solutions. In particular, there is a well-known family of even/odd solutions of period $\pi$ or $2\pi$ (they are \emph{the} Mathieu functions). These solutions are defined by Fourier series, e.g.  
\begin{equation}
ce(z)=\sum_{n\in \mathbb Z} c_n \cos(n z), 
\end{equation}
where one immediately establishes a recursion relation for the $c-$coefficients, 
\begin{equation}
V_{2n}\, c_{2n}=c_{2n+2}+c_{2n-2}, \qquad V_{2n}=[a+(2n)^2]/q,
\end{equation}
which lead to a continuous-fraction relation for the ratios $G_{2n}=c_{2n+2}/c_{2n}$, together with the first ratio $G_0=a/q$. In turn this establishes a family of curves in the $(a,q)$ which are allowed for this type of (even-, $\pi$-periodic) solution. Besides the periodic solutions, there are also the so-familiar Bloch solutions (in solid state these correspond to ``electrons moving through the crystal''), 
\begin{equation}
\psi(z)=e^{ipz} \sum_{n\in \mathbb Z} \tilde c_n \cos(n z),
\end{equation}
where we have used the tilde to emphasize, that a new set of $c_n$ needs to be obtained for each quasi-momentum $p$. The solution to the double eigenvalue problem (family of curves in the (a-q) plane) is provided from two sets of continued fractions, $G_m=c_{m+2}/c_{m}$, $H_m=c_{m-2}/c_m$ fulfilling the condition $G_0H_0=1$. 

A further large family of almost distinguished solutions is provided (see \cite{Abramowitz}, \S 20.4) by modi\-fications of the Floquet (Bloch) solutions, with the \emph{same set of coefficients $\tilde c_m$}, multiplying the functions
\begin{equation}
\vp_k(z)= (-x)^k\, \Be_{k}(2\sqrt{K^2}x), \qquad x= \left[\frac{\cos(z-b)}{\cos(z+b)}\right]^{1/2}.
\end{equation}
The coefficient $b$ in the independent variable, as well as the particular type of the Bessel function, may be chosen freely. These new types of functions, with different behavior in the complex $z$ plane, has, due to the recursion relations, exactly the same spectrum in the (a-q) plane. 

Still further solutions, employing \emph{the same coefficients $\t c_m$} are provided by
\begin{equation}
y(z)=\frac{1}{\t c_{2s}}\, \sum_{n\in Z} \t c_{2n} (-)^n \Be_{n+p +s}(\sqrt{K^2}e^{iz}) J_{n-s}(\sqrt{K^2}e^{-iz}). 
\end{equation}
with an arbitrarily chosen $s$, arbitrary Bessel function $\Be$, and the Floquet (Bloch) momentum $p$.

If, for a given boundary condition, some of the solutions of the above type can be shown to have the right behavior at the boundary, than the whole spectrum (family of lines in $(K^2,M^2)$ plane) \emph{is identical} to the corresponding Floquet (Bloch) spectrum (with the index $p$). This would be so, because the spectrum is determined by the coefficients $\t c_n$, and they, in turn, depend only on $p$ (and the type of Floquet solution used). However, at this point none of these standard solutions of the Mathieu equation appears to play a role for the vortex-bound-state problem. On the other hand standard approximative methods associated with the theory of Mathieu equations \cite{Abramowitz,BenderOrszag} lead to efficient approximations for the positions of bound states and for the scattering characteristics, in particular in the case of large $M^2$.

\section{Note about the  employed numerical method}
The Mathieu-nature of the sole non-trivial differential equation of this paper makes the use of numerical methods unnecessary. However, it proved more efficient to actually use them when dealing with the equation (due to the boundary conditions in a form not standard for Mathieu problems, and due to the robustness of numerical methods for solving ODEs). The radial equation has therefore been implemented in Mathematica, and solutions for arbitrary $\w,p,m$ were obtained starting from the appropriate boundary condition at small $r$. Positions of bound states were obtained by searching for zeroes of the Wronskian of the numerical solution with the asymptotic solution which is damped for $r\ra\infty$ (functions of Bessel-K type, $K_M(Kx)$, see \ref{general_con}).


\bibliography{reference}

\end{document}